\documentclass[bibyear]{aa} 

\usepackage{natbib}
\bibpunct{(}{)}{;}{a}{}{,} 
\usepackage{graphicx}
\usepackage{sidecap}
\usepackage{txfonts}

\usepackage{xcolor}

\begin{document} 

   \title{Multiwavelength observations of KS 1947+300}

   \subtitle{}

   \author{Wei Liu \inst{1}
          \and
          Pablo Reig \inst{3,4}
          \and
          Jingzhi Yan \inst{1} 
          \and
          Peng Zhang \inst{5,6}
          \and
          Xiukun Li \inst{1,2}
          \and
          Bo Gao \inst{1,2}
          \and
          Guangcheng Xiao \inst{1}
          \and
          Qingzhong Liu \inst{1}
          }

   \institute{Key Laboratory of Dark Matter and Space Astronomy, Purple Mountain Observatory, Chinese Academy of Sciences, \\Nanjing, 210023, P.R. China
         \and
             School of Astronomy and Space Science, University of Science and Technology of China, Hefei, 230026, P.R. China
        \and
            Institute of Astrophysics, Foundation for Research and Technology-Hellas, 71110 Heraklion, Greece
        \and
            Physics Department, University of Crete, 71003 Heraklion, Greece
        \and
            College of Science, China Three Gorges University, Yichang 443002, China
        \and
            Center for Astronomy and Space Sciences, China Three Gorges University, Yichang 443002, China\\
            \\
            \email{a010110288@163.com}, {jzyan@pmo.ac.cn}, {zhangpeng@ctgu.edu.cn}
            }

\date{Received <date> /
       Accepted <date>}
       

  \abstract
    {KS 1947+300 is a Be/X-ray binary. Despite its nearly circular orbit, it displays both giant and regular less intense X-ray outbursts. According to the viscous decretion disk model, such low eccentric binaries should not show periodic outbursts.}
    {We have studied the long-term optical variability of KS 1947+300 and its relationship with X-ray activity. Our objective is to investigate the origin of this variability.  }
    {We have analyzed data covering more than 20 years of observations. In the optical band, we have analyzed spectra and light curves. We measured the strength of the H$\alpha$ and He I 6678 \AA\ lines. In the X-ray band, we studied the long-term light curves provided by several all-sky monitors. 
    }
    {KS 1947+300 exhibits changes in brightness and H$\alpha$ emission on time scales from months to years.
    The optical and IR variability shows small amplitude changes during the active X-ray state, and a long, smooth decrease during the quiescent state. The fact that the amplitude of variability increased with wavelength suggests that the long-term decrease of the optical emission is due to the weakening of the circumstellar disk.
    Structural changes in the disk may also be the origin of the periodic signals with periods $\sim200$ days detected in the ZTF-{\it g} and {\it r} band light curves. 
    We speculate that these changes are related to the mechanism that ejects matter from the photosphere of the Be star into the disk. We have also studied the X-ray variability that manifested as a series of type I outbursts after the two giant outburst in 2000 and 2013 and found that the intensity and peak orbital phase differ from outburst to outburst.  The type I outbursts in KS 1947+300 are not strictly periodic. This irregularity could result from precession of the interacting points between the neutron star and the disk, namely the disk apastron and the two nodes of the disk.
    }
    {The long-term changes in optical continuum and line emission and the X-ray variability patterns are attributed to an evolving and distorted decretion disk. }


   \maketitle

\section{Introduction}\label{sec:ks_Introduction}

KS 1947+300 was discovered by the \emph{Kvant}/TTM coded-mask X-ray spectrometer on the \emph{Mir} space station on June 8, 1989. At that time, the source reached a peak flux of 70 $\pm$ 10 mCrab in the 2--27 keV band; The 1989 giant outburst remained active for 35 days before its X-ray flux dropped below the instrument's detection limit of 10 mCrab \citep{1990SvAL...16..345B}. Studies conducted by \citet{1991SvAL...17..415G} and \citet{1991SvAL...17..399G} revealed that the spectra of the optical counterpart to KS 1947+300 exhibited the H$\alpha$ line in emission, defining KS 1947+300 as a Be/X-ray binary (BeXB).
In April 1994, another giant outburst of KS 1947+300 reached a peak flux of 50 mCrab in the 20--75 keV band, lasting for 33 days \citep{1995ApJ...446..826C}.
The system was again bright in X-rays at the end of 2000; In this occasion, the outburst reached a peak flux of 120 mCrab (\emph{RXTE}/ASM, 2--10 keV) and lasted for nearly 150 days \citep{2003A&A...397..739N}.
The latest giant X-ray outburst occurred in 2013 with a peak flux of 140 mCrab in the 2--20 keV energy band (\emph{MAXI}) and 320 mCrab in the 15--50 keV band (\emph{Swift}-BAT), making it one of the largest giant outburst ever recorded for KS 1947+300. The giant (or type II) X-ray outbursts in December 2000 and October 2013 were followed by a series of normal (type I) outbursts. X-ray pulsations with a pulse period of 18.7579 $\pm$ 0.0005 s were detected by 
\citet{2000IAUC.7523....2L} and \citet{2000IAUC.7531....4S}  during the normal outbursts in November 2000.

BeXBs are typically divided into persistent and transient sources based on their X-ray long-term variability \citep{2011Ap&SS.332....1R}. Transient BeXBs display two types of X-ray outbursts: type I (or normal) outbursts and type II (or giant) outbursts.
Type I outbursts are regular and (quasi)periodic outbursts, which normally peak at or close to periastron passage of the neutron star, and reach maximum luminosities $L_{x} \leq 10^{37}$ erg s$^{-1}$. They tend to cover a relatively small fraction of the orbital period (typically 0.2--0.3 $P_{orb}$). 
While, type II outbursts reach luminosities of the order of the Eddington luminosity for a neutron star, $L_{X} \backsim 10^{38}$ erg s$^{-1}$.  Unlike normal outbursts, giant outbursts have no consistently preferred orbital phase and last for a large fraction of an orbital period or even for several orbital periods.


\citet{2004ApJ...613.1164G} obtained an orbital solution for KS 1947+300 with the following orbital parameters: projected semi-major axis $a \times sin{\it i} = 137 \pm 3$ lt-s , orbital period $P_{\rm orb} = 40.415\pm 0.010$ d and orbital eccentricity $e = 0.033\pm0.013$.
\citet{2005AstL...31...88T} estimated the magnetic field strength of the neutron star to be $2.5 \times 10^{13}$ G, based on the correlation of the pulse period derivative and the X-ray luminosity. However, this value of the magnetic field is larger than typical of accreting pulsars ($B\leq 10^{12}$ G) and would imply an unrealistic cyclotron resonant scattering feature at $\sim$ 220  keV. In fact, a cyclotron line at 12.5 keV (i.e. $B \sim 1.1 \times 10^{12}$ G) was reported by \citet{2014ApJ...784L..40F}. However, \citet{2020MNRAS.493.3442D} invoked the low-significance of the residuals and calibration issues to question the reality of that feature. Hence the actual value of the magnetic field remains open.


KS 1947+300 has been observed by many X-ray missions and its X-ray variability has been studied in detail. In contrast, the number of works that study the properties of this system in the optical band reduce to two. \citet{2003A&A...397..739N} presented the first optical spectroscopy and optical and infrared photometry of the counterpart to the transient X-ray source KS 1947+300. They showed it to be a moderately reddened V = 14.2 early-type Be star located in an area of low interstellar absorption slightly above the Galactic plane. They derived a spectral type B0Ve and estimated a distance $\sim$10 kpc. \citet{2007AN....328..142K} performed optical photometry of KS 1947+300 with high cadence and found an increase of about 0.1 mag in coincidence with and increase in the X-ray flux from quiescence to $5 \times 10^{36}$ erg s$^{-1}$, typical behavior of a type I outburst. Other works that report optical observations of KS 1947+300 in the context of a global study of the optical properties of BeXBs are given in \citet{2015A&A...574A..33R}, \citet{2016A&A...590A.122R}, and \citet{2022A&A...667A..18R}.

In this article, we present the most comprehensive optical analysis of KS 1947+300, including spectroscopic and photometric data from various bands: the infrared bands at 3.4 $\mu$m and 4.6 $\mu$m, the $V$-band, the H$\alpha$ line, and the He I $\lambda6678$ line. Additionally, we provide the X-ray light curve and spin variations of the neutron star for reference. These observations reveal the gradual decrease in the size of the circumstellar disk around the Be star following giant X-ray outbursts. Our main focus is on the changes in the optical band and their connection with the evolution of the circumstellar disk of the Be star.


\section{Observations and data analysis}\label{sec:ks_Observation}

\subsection{Optical spectroscopy} 

Optical spectroscopic observations were  obtained from three telescopes at three different observatories: The observations from the Xinglong Station of National Astronomical Observatories in Hebei province (China) were obtained with the spectrometer OptoMechanics Research (OMR) or BAO Faint Object Spectrograph and Camera (BFOSC) on the 2.16 m telescope; while the observations from the Lijiang station of Yunnan Astronomical Observatory in Yunnan province (China) used the spectrometer Yunnan Faint Object Spectrograph and Camera (YFOSC) on the 2.4 m telescope.
The OMR is equipped with a 1024 $\times$ 1024 (24 micron) pixels TK1024AB2 CCD. The OMR Grism 4 is 1200 lines $\rm mm^{-1}$, giving a nominal dispersion of 1.0 \AA\ pixel$^{-1}$ and a spectral resolution of 2.7 \AA. The average resolving power is $R\sim2300$ over the wavelength range of 5500--6900 \AA\ \citep{2016PASP..128k5005F}. 
The BFOSC is equipped with a 2048 $\times$ 2048 (15 micron) pixels Loral Lick 3 CCD. The nominal dispersion of the BFOSC Grism 8 is 1.79 \AA\ pixel$^{-1}$, covering the wavelength range of 5800--8280 \AA\ \citep{2016PASP..128k5005F}.  The spectral resolution is about 2.4 \AA\ ($R\sim 3000$).
The YFOSC is equipped with a 2k $\times$ 4k (13.5 micron) pixels E2V 42-90 CCD. The nominal dispersion of the YFOSC Grism 8 is 1.47 \AA\ pixel$^{-1}$. The spectral resolution is about 10.3 \AA\, covering the wavelength range of 4970--9830 \AA. The average spectral resolving power is about 700.
In addition, we analyzed optical spectroscopic observations obtained from the 1.3 m telescope of the Skinakas observatory (SKO) in Crete (Greece). The 1.3 m telescope was equipped with a 2048 $\times$ 2048 (13.5 micron) pixels ANDOR IKON CCD, a 1302 lines $\rm mm^{-1}$ grating, and a slit with of 160 $\nu$m giving a nominal dispersion of $\sim$\,0.94 \AA\ pixel$^{-1}$. The spectral resolving power of the 1.3 m telescope is about 1300, covering the wavelength range of 5400--7300 \AA. Data prior to 2016 were taken from \citet{2016A&A...590A.122R}. 

We used the Image Reduction and Analysis Facility (IRAF)\footnote{IRAF is distributed by NOAO, which is operated by the Association of Universities for Research in Astronomy, Inc., under cooperation with the National Science Foundation.} software package to reduce and analyze all the spectra, including bias-subtracted correction and flat-field correction, and cosmic-ray subtraction. A Helium-Argon calibration lamp was employed to obtain the pixel--wavelength relationship. In order to ensure the consistency of spectral processing, all spectra were normalized to adjacent continua. We measured the equivalent widths of the H$\alpha$ lines (hereafter EW(H$\alpha$) for short) for five times, each measurement with a different selection of the continuum. The final EW(H$\alpha$) is the average of the five measurements, and the error is the standard deviation. The typical error of EW(H$\alpha$) is within 5\%. The value of the error is determined by the quality of the continuum. The equivalent widths of the He I $\lambda$6678 lines (hereafter EW(He I $\lambda$6678) for short) were obtained following the same method as EW(H$\alpha$).

The log of the spectroscopic observations is given in Tables~\ref{table_ks_spec_2m}, \ref{table_ks_spec_1.3m_1} \& \ref{table_ks_spec_1.3m_2}. EW(He I $\lambda$6678) and EW(H$\alpha$) are plotted in the third and sixth panels of Fig.~\ref{ks_multiplot}, respectively. The evolution of the EW(H$\alpha$) is plotted in the third panel of of Fig.~\ref{ks_longterm}. The evolution of the H$\alpha$ line profiles from 2018 to 2022 is plotted in Fig.~\ref{ks1_profile_Ha_peak_valley}.


\subsection{Optical photometry}

Optical photometric observations were obtained from five telescopes at four different observatories: From the Xinglong Station of National Astronomical Observatories, Chinese Academy\\ of Sciences (NAOC), observations were obtained with the Tsinghua-NAOC Telescope (TNT, 80 cm) and the 60 cm telescope; from the Lijiang station of Yunnan Observatories (YNAO), the data came from the 2.4 m telescope; from the Yaoan astronomical observation station of Purple Mountain Observatory (PMO), the data came from the Yaoan High Precision Telescope (YAHPT); from the Skinakas observatory, the data were obtained with the 1.3 m telescope.

The TNT (80 cm) is an equatorial-mounted Cassegrain system with a focal ratio of f/10, made by AstroOptik, funded by Tsinghua University in 2004 and jointly operated with NAOC, which is equipped with the PI VersArray 1300B LN 1340 $\times$ 1300 thin, back-illuminated CCD with a 20 $\mu$m pixel size \citep{2012RAA....12.1585H}. In this configuration, the plate scale is 0.52" pixel$^{-1}$ and gives a field of view of $11.5 \times 11.2$ $\rm arcmin^{2}$.
The 60 cm telescope is an equatorial-mounted system with a focal ratio of f/4.23, which is equipped with the Andor DU934P-BEX2-DD 1024 $\times$ 1024 CCD and provides a field of view of 18 $\times$ 18 $\rm arcmin^{2}$.
The 2.4 m telescope is an altazimuth-mounted Cassegrain system with a focal ratio of f/8,  which is equipped with the E2V CCD42-90  2k $\times$ 2k thin, back-illuminated, deep-depletion CCD with a 13.5 $\mu$m pixel size. In this configuration, the plate scale is 0.28" pixel$^{-1}$ and gives a field of view of 9.6 $\times$ 9.6 $\rm arcmin^{2}$.
The YAHPT (80 cm') is an altazimuth-mounted, RC optical system with a focal ratio of f/10, made by Astro Systeme Austria, which is equipped with the PIXIS 2048B back-illuminated CCD with a 13.5 $\mu$m pixel size.  In this configuration, the plate scale is 0.347" pixel$^{-1}$, providing a field of view of 11.8 $\times$ 11.8 $\rm arcmin^{2}$.

Three different CCDs we used during the observations from the Skinkas observatory: Before June 2007, a 1024 × 1024 SITe chip with a 24 $\mu$m pixel size (corresponding to 0.5 arcsec on the sky) was used. From July 2007 to October 2017, the telescope 
was equipped with an ANDOR CCD DZ436. From 2018, an IKON-L 926 was installed. Both have an array of 2048 × 2048 13.5 $\mu$m pixel size (corresponding to 0.28 arcsec on sky), hence providing a field of view of 9.5 square arcmin.

In all five telescopes, KS 1947+300 was observed through the standard Johnson-Cousins $B$, $V$, $R$, and $I$ filters. The photometric data reduction was performed using standard routines and aperture photometry packages (some from the zphot package) in IRAF, including bias subtraction and flat-field correction. In order to derive the variation in the optical brightness, we selected the reference star C1 ($\alpha$: 19 49 41.5, $\delta$: +30 13 47, J2000) \cite[according to ][ the average magnitudes of the reference star are B = 15.329 ± 0.006, V = 14.504 ± 0.006, R = 14.002 ± 0.006, and I = 13.537 ± 0.009]{2015A&A...574A..33R} in the field of view of KS 1947+300 to derive its differential magnitudes. In the case of the Skinakas observations, the quoted magnitudes resulted from absolute photometry.
Standard stars from the Landolt's list \citep{2009AJ....137.4186L} were used for the transformation equations. Observations prior to 2015 were taken from \cite{2015A&A...574A..33R}. Observation after that year represent new data. The photometric magnitudes are given in Tables~\ref{table_ks_phot_0}, \ref{table_ks_phot_1} \& \ref{table_ks_phot_2}.

The long-term $R$-band light curve is plotted in the second panel of Fig.~\ref{ks_longterm}. It covers the period 2000-2023, while the Johnson $V$-band light curve covering the period after the 2013 X-ray outburst is plotted in the fourth panel of Fig.~\ref{ks_multiplot}. The evolution of the $(B-V)$ color index is plotted in the fourth panel of Fig.~\ref{ks_longterm} and the seventh panel of Fig.~\ref{ks_multiplot}.



\subsection{ZTF data}

The ZTF Observing System is installed on the 48 inch Samuel Oschin Telescope (Schmidt-type) at the Palomar Observatory. It uses 16 e2v 6k x 6k CCD231-C6 CCDs. The 15 microns pixels provide a pixel scale of 1"/pixel and a field of view of 47 square degrees. With this large field of view, ZTF scans the entire Northern sky every two days. ZTF uses three custom made filters, ZTF-g, ZTF-r, and ZTF-i, centered at around 480 nm, 650 nm, and 800 nm, respectively \citep{2019PASP..131a8002B}.

We downloaded the KS 1947+300 light curves from the Infrared Science Archive (IRSA) \footnote{https://irsa.ipac.caltech.edu/Missions/ztf.html}. The data correspond to the ZTF public Data Release (DR) 18 (release on July 07, 2023). The ZTF-{\it r} and {\it g} light curves are shown in Figs.~ \ref{ks_longterm} and \ref{ks_multiplot}.


\subsection{\emph{NEOWISE} data}

We made use of the light curves in the W1 (3.4 $\mu$m) and W2 (4.6 $\mu$m) bands provided by \emph{NEOWISE} \citep{2011ApJ...731...53M} project through the IRSA viewer\footnote{https://irsa.ipac.caltech.edu/irsaviewer}, and plot them in the fifth panel of Fig.~\ref{ks_multiplot}.

\subsection{X-Ray data}

\begin{figure*}
   \centering
   \includegraphics[bb=-300 200 1600 2000,width=1.0\textwidth]{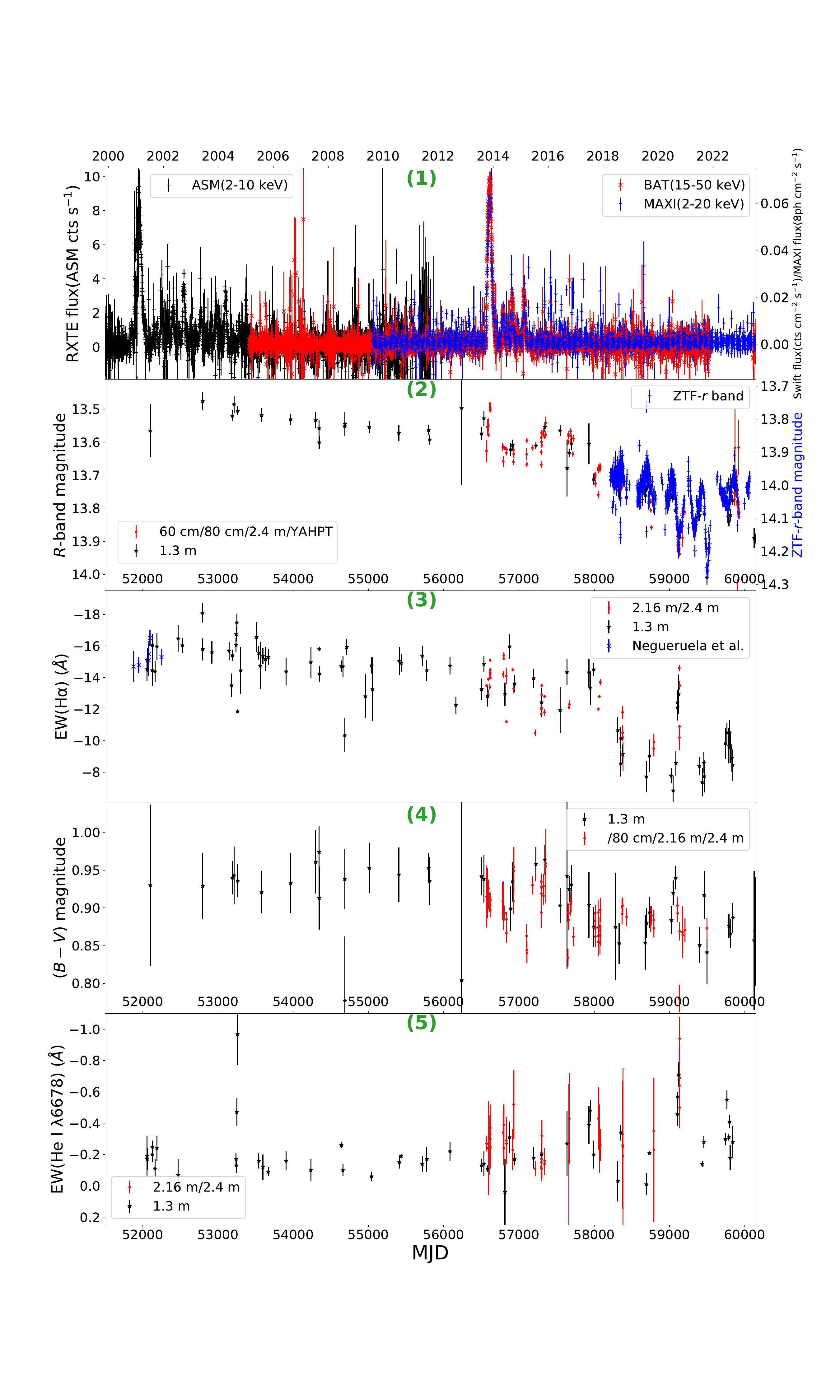}
   \caption{\,Panel 1: X-ray light curves. \emph{RXTE}-ASM (2-10 keV), \emph{Swift}-BAT (15–50 keV), and \emph{MAXI} (2–20 keV) are separately labeled with black plus signs, red crosses, and blue downward triangles. The position of the first day of each year is also indicated above panel 1.
   Panel 2: Long-term light curve of $R$-band. Our data from 60 cm, 80 cm, 2.4 m telescopes, and YAHPT are labeled with red dots. Our data from 1.3 m telescope are labeled with black stars, and the data before 2015 are from \citet{2015A&A...574A..33R}. The data from ZTF-{\it r} band are labeled with blue downward triangles.
   Panel 3: Long-term variation of EW(H$\alpha$). Our data from 2.16 m and 2.4 m telescopes are labeled with red dots. Our data from 1.3 m telescope are labeled with black stars, and the data before 2016 are from \citet{2016A&A...590A.122R}. And the data from \citet{2003A&A...397..739N} are labeled with blue crosses.
   Panel 4: Long-term variation of $(B-V)$ color index. Our data from 80 cm and 2.4 m telescopes, and YAHPT are labeled with red dots. Our data from 1.3 m telescope are labeled with black stars, and the data before 2015 are from \citet{2015A&A...574A..33R}.
   Panel 5: Long-term equivalent widths of He I $\lambda$6678 line. Our data from 2.16 m and 2.4 m telescopes are labeled with red dots. Our data from 1.3 m telescope are labeled with black stars.
   \label{ks_longterm}
 }
\end{figure*}

\begin{figure*}
\centering
\includegraphics[bb=-410 300 1710 2700,width=1.0\textwidth]{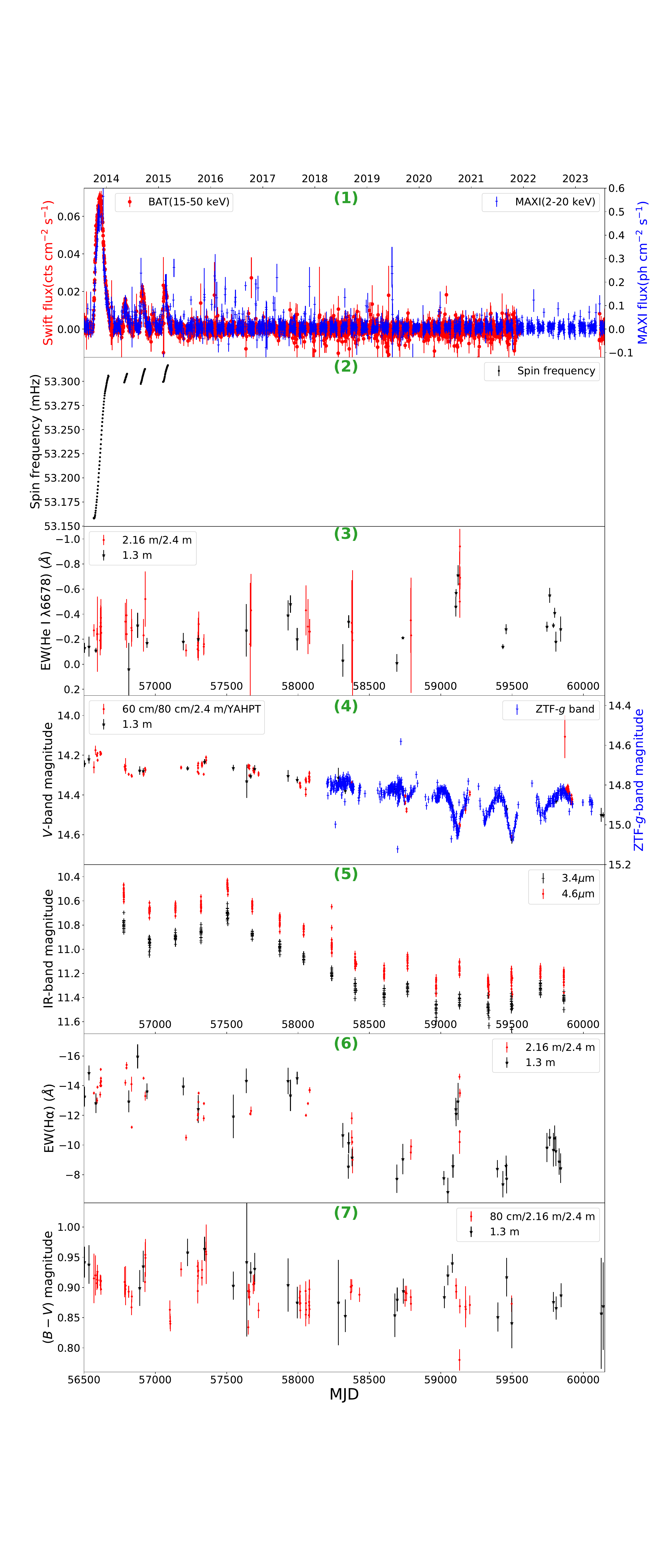}
\caption{\,Panel 1: X-ray light curves. Data from \emph{Swift}-BAT (15--50 keV) and \emph{MAXI} (2--20 keV) are separately labeled with red dots and blue downward triangles. The position of the first day of each year is also indicated above panel 1.
Panel 2: Barycentered and orbit-corrected spin-frequency history measured with \emph{Fermi}-GBM.
Panel 3: Equivalent widths of He I $\lambda$6678 line. Our data from 2.16 m and 2.4 m telescopes are labeled with red dots. Our data from 1.3 m telescope are labeled with black stars.
Panel 4: Light curves of optical $V$-band. Our data from 60 cm, 80 cm, 2.4 m telescopes, and YAHPT are labeled with red dots. Our data from 1.3 m telescope are labeled with black stars, and the data before 2015 are from \citet{2015A&A...574A..33R}. The data from ZTF-{\it g} band are labeled with blue downward triangles.
Panel 5: Light curves of infrared 3.4 $\mu$m and 4.6 $\mu$m bands from \emph{NEOWISE}. They are separately labeled with black plus signs and red dots, respectively.
Panel 6: Variation of EW(H$\alpha$). Our data from 2.16 m and 2.4 m telescopes are labeled with red dots. Our data from 1.3 m telescope are labeled with black stars, and the data before 2016 are from \citet{2016A&A...590A.122R}.
Panel 7: Variation of $(B-V)$ color index. Our data from 80 cm and 2.4 m telescopes, and YAHPT are labeled with red dots. Our data from 1.3 m telescope are labeled with black stars, and the data before 2015 are from \citet{2015A&A...574A..33R}.
\label{ks_multiplot}}
\end{figure*}

The All-Sky Monitor (ASM)\footnote{ASM consists of three wide-angle shadow cameras equipped with proportional counters with a total collecting area of 90 square cm.}\footnote{https://heasarc.gsfc.nasa.gov/FTP/xte/data/archive/ASMProducts/\\definitive\_1dwell/lightcurves/xa\_ks1947+300\_d1.lc} on board \emph{RXTE}, BAT \footnote{https://swift.gsfc.nasa.gov/results/transients/weak/KS1947p300/} on board \emph{Swift} (Krimm et al. \citeyear{2013ApJS..209...14K}), \emph{MAXI}\footnote{http://maxi.riken.jp/star\_data/J1949+302/J1949+302.html}, and the Gamma-ray Burst Monitor (GBM)\footnote{https://gammaray.nsstc.nasa.gov/gbm/science/pulsars/lightcurves/\\ks1947.html} on board \emph{Fermi} \citep{2009ApJ...702..791M} have been monitoring KS 1947+300 in the hard X-ray energy band (2--10 keV with ASM, 15--50 keV with BAT, 2--20 keV with \emph{MAXI}, and 12--50 keV with GBM). Two giant X-ray outbursts and several normal outbursts were detected between 2000 and 2023. The X-ray band light curves from ASM (2--10 keV), BAT (15–50 keV), and \emph{MAXI} (2--20 keV) are plotted in the first panel of Figs.~\ref{ks_longterm} and \ref{ks_multiplot}. The spin-frequency history measured by GBM is plotted in the second panel of Fig.~\ref{ks_multiplot}.

\section{Results}\label{sec:ks_Result}

The profile changes of emission lines in Be stars, especially of the H$\alpha$ line, can be used to track the dynamic evolution of the Be circumstellar disk \citep{2001A&A...369..117N}. Unfortunately, the spectral resolution of most of our spectra is too low to perform a detailed analysis, but it is still possible to obtain some information from intensity changes in the continuum and line emission.

\subsection{Long-term optical variability}
\label{long-term}


The X-ray variability was dominated by two giant (type II) outbursts that took place in 2000 and 2013. They were followed by a series of normal (type I) outbursts, whose peak intensities were roughly modulated by the orbital period. The type I outburst phase was particularly intense from November 2000 to January 2005 and from February 2014 to March 2015. In the optical band, the behavior of the continuum and line emission during the type I active phases in X-rays is similar for the two intervals, although the source was in a fainter state after the 2013 outburst. The optical variability is characterized by a phase of relatively stable emission after the giant outbursts that coincides with the presence of type I outbursts, followed by a decrease in the overall brightness once the X-ray activity ceased (see Fig.~\ref{ks_longterm}).  This decrease in brightness after the 2013 outburst is more distinctly observed in the $R$ and $I$ bands than in the $B$ and $V$ bands: on average the amplitude of variation amounted to 0.15, 0.2, 0.24, and 0.26 mag for the $B$, $V$, $R$, and $I$, respectively. During the same interval, EW(H$\alpha$) decreased by $\sim6$ \AA.



Figure~\ref{ks_multiplot} focuses on the variability after the 2013 major outburst, where we replaced the $R$ band with the $V$ band, and added the neutron star spin-frequency history and \emph{NEOWISE} 3.4 $\mu$m and 4.6 $\mu$m infrared band light curves. 

In the X-ray band, KS 1947+300 displayed one giant outburst and five normal outbursts from 2013 to 2015, and then it went into a quiescent state. The first, third, and fifth outbursts (number 18, 20, and 22 in Fig.~\ref{typeI_BAT}) were significantly brighter than the second and forth ones (number 19 and 21 in Fig.~\ref{typeI_BAT}). The changes in the neutron star's spin frequency are very interesting. During the giant outburst, there is a distinct increase in the spin frequency. The amount of accreted matter transferred enough angular momentum to the neutron star to increase its rotation speed. The spin period decreased from 18.81 s when the outbursts started to 18.78 s at the peak of the outburst. This value appears as some kind of equilibrium period because as soon as the accretion of matter stopped, the spin frequency decreased, while it increased again when the accretion rate was large (i.e. during the three brightest type I outbursts).

The IR variability was similar to the optical one, namely small amplitude changes during the active X-ray state, and a long, smooth decrease during the quiescent state. 

The He I line at 6678 \AA\ appeared most of the time in emission with its EW(He I $\lambda6678$)  stable at $-$0.3 \AA, with small variations.  Around October 2020, an abrupt increase was measured in coincidence with a similar increase in EW(H$\alpha$).

\subsection{Evolution of H$\alpha$ line profiles}\label{sec:ks_Result_Halpha}


\begin{figure}
   \centering
   \includegraphics[width=\hsize]{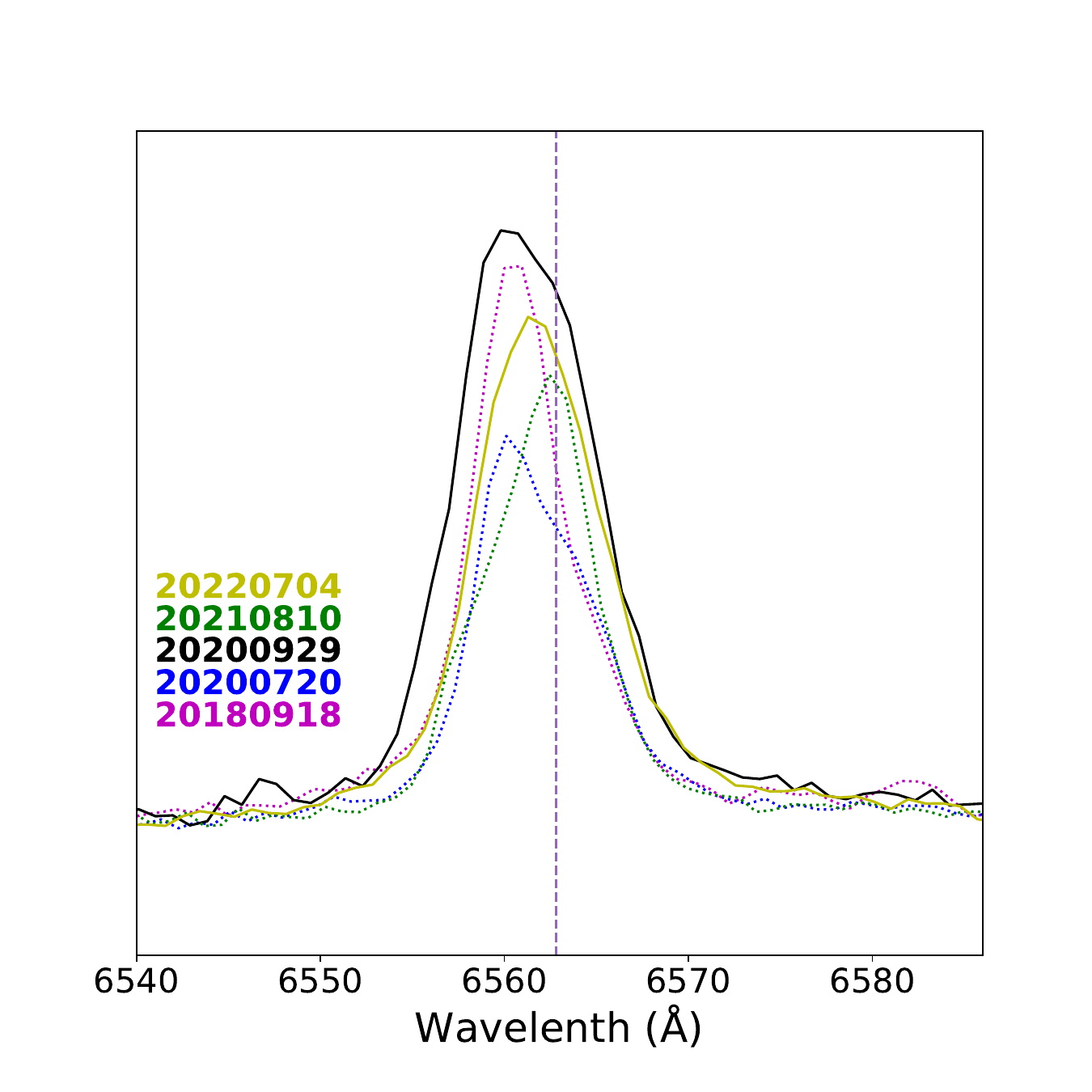}
      \caption{\, Evolution of H$\alpha$ line profiles. The vertical lines mark the rest wavelength of the H$\alpha$ line. The dotted line profiles were obtained when the optical brightness was near peak values, while the solid line profiles were obtained when the optical brightness was near valley values.
              }
         \label{ks1_profile_Ha_peak_valley}
   \end{figure}

The $H\alpha$ line always appears in emission with a broad, sometimes distorted, single peak profile. The low spectral resolution prevents us from a detailed analysis but there is some evidence of V/R variability, which might be the origin of the asymmetry observed in some profiles (i.e. the profile is affected by the precession of a density perturbation in the disk; when it approaches the observer, the intensity increases in the blue part of the profile, while when it moves away from the observer, it affects the red part of the line) \citep{1994A&A...288..558T}. Figure~\ref{ks1_profile_Ha_peak_valley} shows some characteristic profiles of the $H\alpha$ line. These spectra were selected to coincide with the peaks and troughs of the photometric quasiperiodic variability (see Sect.~\ref{sec:ks_Discu_periodic_signals_in_ZTF}). However, no significant difference is observed.

\section{Discussion}\label{sec:ks_Discu}

\subsection{Changes of circumstellar disk state}

Previous works claimed that KS 1947+300 is a Be/X-ray binary whose circumstellar disk's scale is not sensitive to X-ray outbursts. That is to say, the occurrence of X-ray outbursts does not lead to significant changes in the H$\alpha$ line intensity or shape. 
\citet{2003A&A...397..739N} considered that X-ray outbursts did not have any effect on  EW(H$\alpha$) because of the low inclination of the system, while \citet{2016A&A...590A.122R} believed that the lack of significant change after the giant X-ray outburst should be attributed to the nearly circular orbit of KS 1947+300. The study presented in this work covers a significant longer time span than previous works and clearly reveal that the state of the circumstellar disk changed in response to the X-ray outbursts, resulting in the fading of the optical and infrared brightness and the decrease of EW(H$\alpha$). These changes, however, occur on longer time scales than other well known BeXBs.

The weakening of the line and the fading of the continuum emission started after the 2000 outburst, but accentuated after the 2013 outburst. The decrease in brightness affected more significantly the $R$ and $I$ bands, as well as EW(H$\alpha$). This result clearly indicates that the source of the variability is the decretion disk around the Be star. Therefore, we interpret the long-term decrease of the optical emission as a weakening of the disk.


\subsection{Study of non-periodic type I X-ray outbursts after 2000 and 2013} \label{sec:ks_Discu_non-periodic_Type_I_X-ray_outbursts}

\begin{figure*}
   \centering
   \includegraphics[width=\hsize]{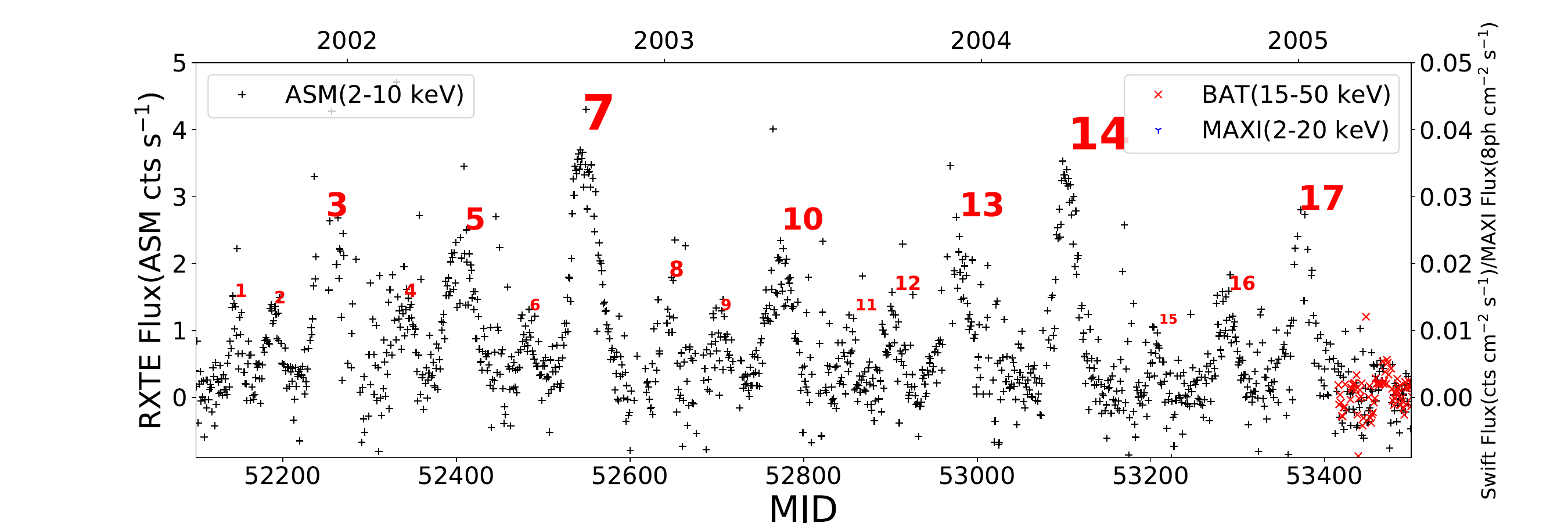}
   \caption{\,Schematic diagram of a series of type I X-ray outbursts after 2000. The font size of the type I outburst numbers represents the peak intensity of each outburst. 
    \label{typeI_RXTE}}
\end{figure*}

\begin{figure*}
   \centering
   \includegraphics[width=\hsize]{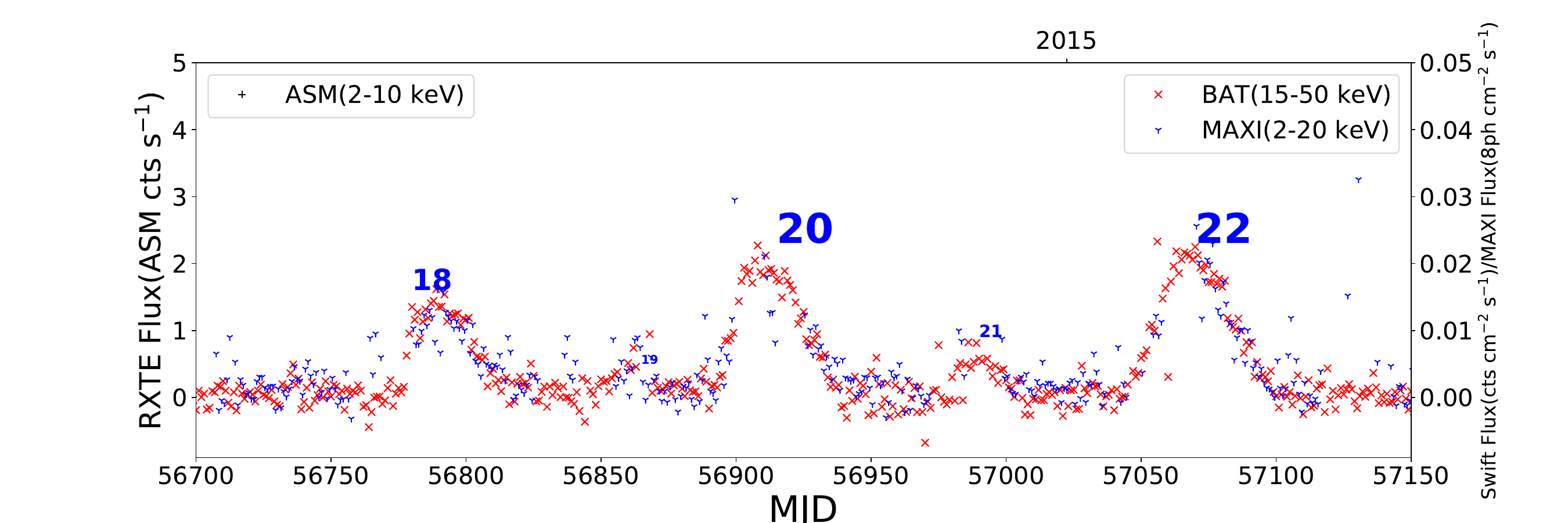}
   \caption{\,Schematic diagram of a series of type I X-ray outbursts after 2013. The font size of the type I outburst numbers represents the peak intensity of each outburst. 
    \label{typeI_BAT}}
\end{figure*}

\begin{figure}
   \centering
   \includegraphics[bb=50 100 700 700,width=\hsize]{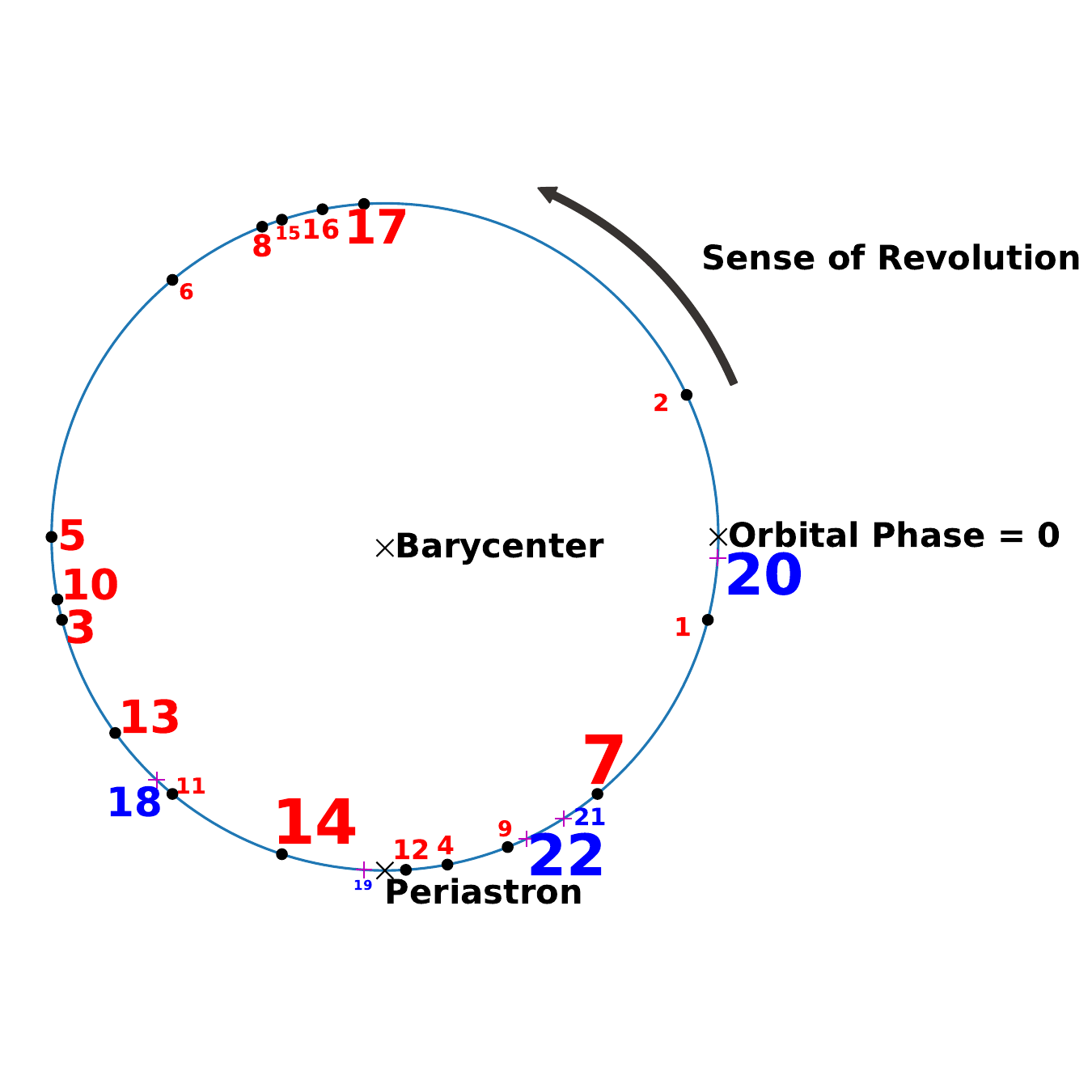}
   \caption{\,Orbital phases diagram of a series of type I X-ray outbursts after 2000 and after 2013. The numbers 1--17 represent the type I outbursts after 2000, while the numbers 18--22 represent the type I outbursts after 2013. The font size of the type I outburst numbers represents the peak intensity of each outburst.
    \label{typeI}}
         
   \end{figure}

\begin{table}
\centering
\caption{Orbital phases (at the peak) of the series of type I X-ray outbursts after 2000 and after 2013. }
\label{table_typeI}
\begin{tabular}{cccc}
\hline\hline
Number &MJD & Phase & Peak Intensity \\
&&& (mCrab) \\
\hline
01 & 52145 & 0.96 & 6.4  \\  
02 & 52190 & 0.07 & 6.0  \\  
03 & 52250 & 0.54 & 17.6 \\
04 & 52340 & 0.78 & 6.4  \\  
05 & 52410 & 0.50 & 10.8 \\
06 & 52485 & 0.36 & 5.6  \\  
07 & 52545 & 0.86 & 17.6 \\
08 & 52645 & 0.31 & 8.0  \\  
09 & 52705 & 0.81 & 5.6  \\  
10 & 52775 & 0.53 & 10.8 \\
11 & 52860 & 0.64 & 5.6  \\  
12 & 52905 & 0.76 & 6.8  \\  
13 & 52980 & 0.60 & 12.0 \\
14 & 53105 & 0.70 & 16.4 \\
15 & 53210 & 0.30 & 4.8  \\  
16 & 53290 & 0.28 & 6.8  \\  
17 & 53370 & 0.26 & 12.4 \\
18 & 56780 & 0.63 & 34.6 \\
19 & 56865 & 0.74 & 10.8 \\
20 & 56915 & 0.99 & 49.8 \\
21 & 56990 & 0.84 & 19.5 \\
22 & 57070 & 0.82 & 49.8 \\
\hline
\end{tabular}
\end{table}

After the two giant X-ray outbursts in 1989 and 1994 \citep{2003A&A...397..739N}, KS 1947+300 became active again in late 2000. After the first weak outburst in November 2000 \citep{2000IAUC.7523....2L,2000IAUC.7531....4S}, KS 1947+300 experienced a giant outburst which lasted about 150 days and reached maximum flux around MJD 51953 (February 13, 2001) at more than 120 mCrab \citep{2008A&A...489..725R}. Subsequently, from the second half of 2001 to the beginning of 2005, the source displayed seventeen normal (type I) outbursts. These outbursts peaked at different intensities and at different time intervals, not exactly coinciding with periastron. 

Figures~\ref{typeI_RXTE} and~\ref{typeI_BAT} show the X-ray activity of KS 1947+300 immediately after the giant outbursts in 2000 and 2013, respectively, where the individual outbursts can be discerned. To aid the reader we have marked each outburst with a number in chronological order.
These type I outbursts did not occur at a fixed orbital phase, as indicated in Table~\ref{table_typeI} and Fig.~\ref{typeI}.
In Fig.~\ref{typeI}, the reference point for the orbital phase is taken at $T_{\rm \pi/2}$ (MJD) = 51985.31, with an orbital period of $P_{\rm orb} = 40.415 \pm 0.010$ d and binary eccentricity 0.033 $\pm$ 0.013 \citep{2004ApJ...613.1164G}.

From Fig.~\ref{typeI}, it can be observed that although there is no clear preference for orbital phases in these two sets of type I outbursts, nearby outbursts often occurred at similar phases. For example, outbursts number 3 \& 5, 6 \& 8, 7 \& 9, 11 \& 13, 12 \& 14, 15 \& 16 \& 17, and 21 \& 22 exhibit such behavior.

The main parameters driving the presence of different type of outbursts (type I versus type II) is believed to be the eccentricity of the binary system and viscosity of the disk because these parameters determine the degree of truncation of the disk by the torques exerted by the neutron star. The viscous decretion disk model \citep{2001A&A...377..161O,2001A&A...369..108N} predicts that in low eccentric orbits the gap between the disk outer radius and the critical lobe radius of the Be star is so wide that, under normal conditions, the neutron star cannot accrete enough gas at periastron passage to show periodic X-ray outbursts (type I outbursts). In contrast, Be/X-ray binary systems with larger orbital eccentricities are expected to display type I X-ray outbursts more easily because the disk truncation occurs at a much higher resonance radius, which is very close to or slightly beyond the critical lobe radius at periastron \citep{2001A&A...377..161O}.

However, a number of low eccentric BeXBs have been reported to show type I outbursts \citep[][and references therein]{2019ApJ...881L..32F}. The most paradigmatic case is KS 1947+300. \citet{2019ApJ...881L..32F} found that the extreme mass ratio of the binary leads to the presence of 3:1 Lindblad resonance inside the Be star disk and this drives eccentricity growth. Hence, even in a circular orbit binary, a nearly coplanar disk around the Be star can become eccentric. As a result, the neutron star is able to capture material every time it passes the disk apastron, thus producing type I outbursts.

Another result that challenges the standard view in BeXBs is the fact that type I outbursts in KS 1947+300 are not periodic. Current belief is that type I outbursts are modulated by the orbital period since they normally occur at or near periastron passage. Therefore, the separation between two such outbursts should correspond with the orbital period within some uncertainty. However, we showed (see Fig~\ref{typeI}) that KS 1947+300 exhibited type I outbursts at different orbital phases. 
An explanation for this behavior was provided by \citet{2021ApJ...922L..37M}. They show that nonperiodic type I outbursts may be temporarily driven in a low eccentricity binary with a disk that is inclined sufficiently to be mildly unstable to Kozai–Lidov oscillations, where the inclination and eccentricity of the disk interchange periodically \citep{2014ApJ...792L..33M,2015ApJ...807...75F}. After several orbital passages of the neutron star, the disk develops an asymmetric structure with one or more spiral arms. Outbursts occur when the neutron star passes close to these spiral arms and pull material from it. The spiral arms may have lower density compared to the circumstellar disk. Therefore, not every accretion event from these arms results in an outburst.
According to \citet{2021ApJ...922L..37M}, a misaligned and eccentric disk can have between one and three outbursts per orbit with varying magnitudes depending on the interacting points between the neutron star and the disk. The three different locations would be the disk apastron and the two nodes of the disk (ascending and descending).
The irregularity on the outbursts would result from variation of the nodal and apsidal precession.

\begin{figure}
   \centering
   \includegraphics[width=\hsize]{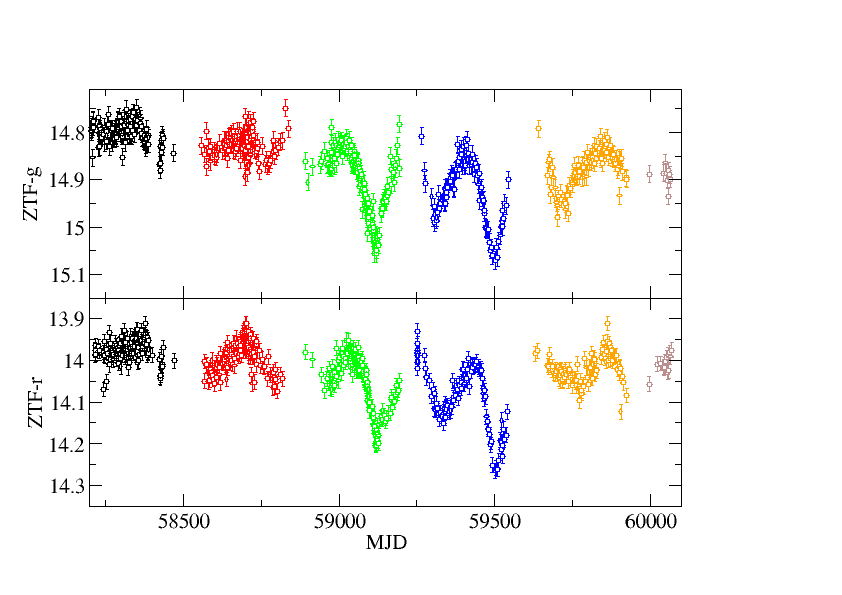}
   \caption{ZTF-{\it g} and ZTF-{\it r} band light curves.}
         \label{ZTF}
   \end{figure}


\begin{figure}
   \centering
   \includegraphics[width=\hsize]{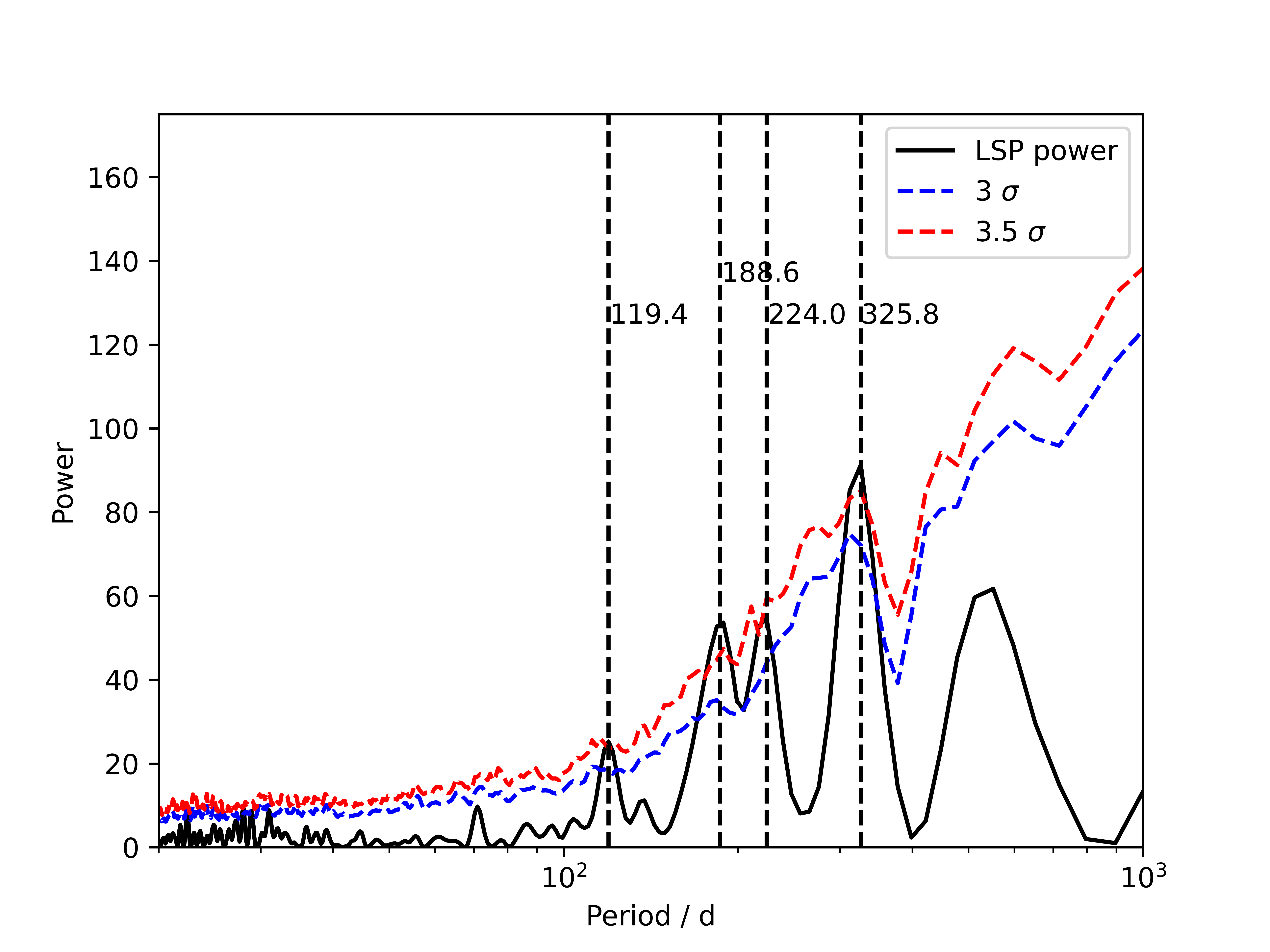}
   \includegraphics[width=\hsize]{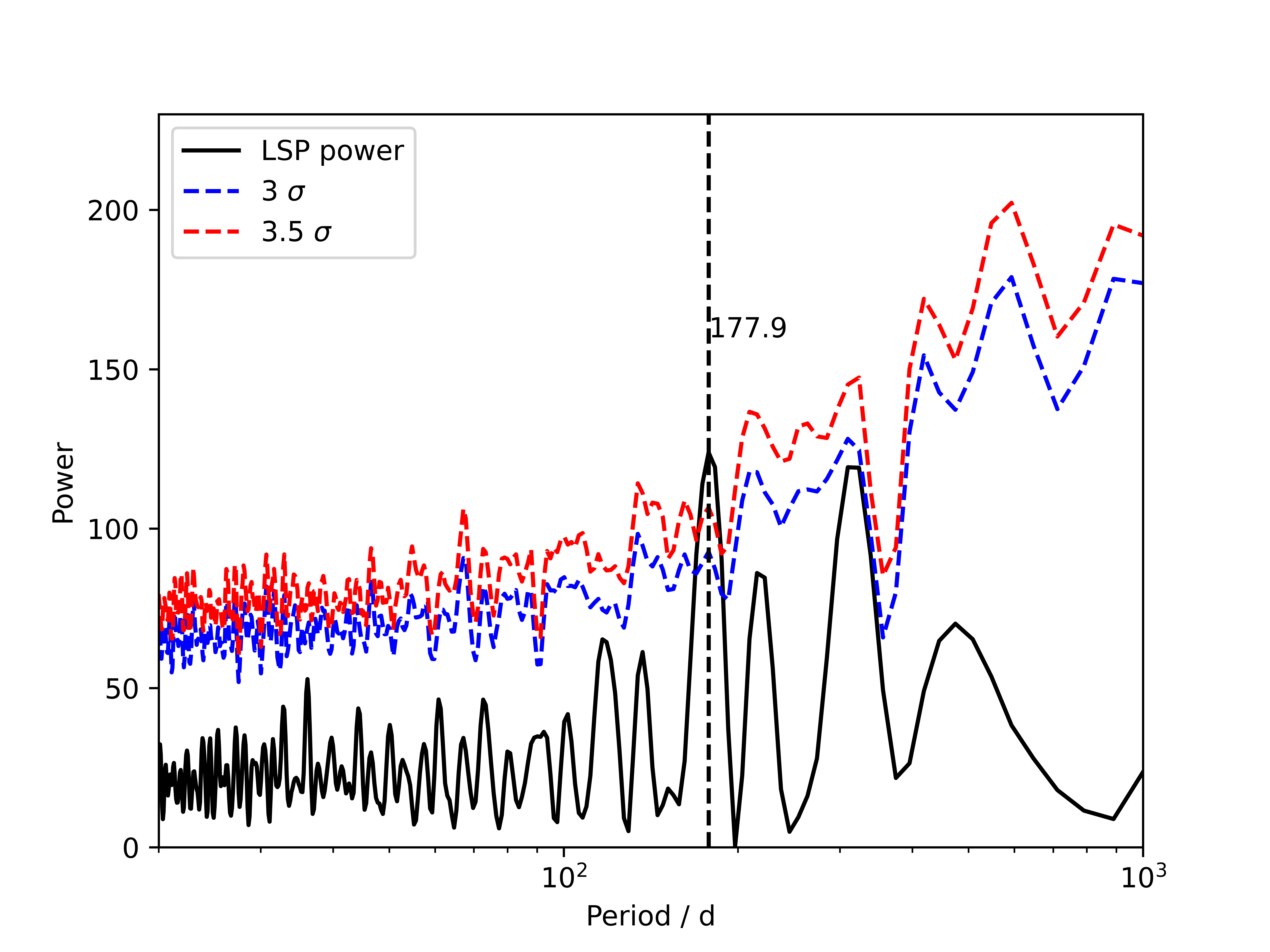}
      \caption{The LSP results for the ZTF-{\it g} (upper panel) and ZTF-{\it r} (lower panel) bands. The black line represents the power values of the LSP. Blue and red dashed lines represent the confidence level curves of 3 $\sigma$ and 3.5 $\sigma$, respectively. The black dashed line indicates signals that exceed the 3.5 sigma threshold.}
         \label{LSP}
   \end{figure}

\subsection{Periodic signals in ZTF-{\it g} and {\it r} band light curves}
\label{sec:ks_Discu_periodic_signals_in_ZTF}

In Sect.~\ref{sec:ks_Result}, we described the overall long-term variability trends (time scales of years). However, an inspection of Fig.~\ref{ks_multiplot} reveals variability in the optical brightness on time scales of months which are particularly apparent when the observation cadence is high (i.e with the ZTF-{\it g} and {\it r} data in the interval 2018-2023). Figure~\ref{ZTF} offers a detailed view of these variations. 

In order to search for possible periodic signals, we employed the method of Lomb-Scargle periodogram (LSP)\citep{1976Ap&SS..39..447L, 1982ApJ...263..835S, 2009A&A...496..577Z}, and estimated the significance of the signals based on their false-alarm probability. We analyzed the data in ZTF-{\it g} and ZTF-{\it r} bands and identified several signals with different periods. To estimate the confidence level more robustly, we generate 2 $\times$ 10$^{4}$ artificial light curves based on the power spectral density (PSD) and the probability density function of the variation of the ZTF-{\it g} and ZTF-{\it r} light curve, respectively \citep{2013MNRAS.433..907E}. The simulated light curves have the full properties of statistics and variability of the ZTF-{\it g} and ZTF-{\it r} light curve. By obtaining the LSP periodograms of these simulated light curves, we derive confidence curves at 3 $\sigma$ and 3.5 $\sigma$ levels. 

Figure~\ref{LSP} shows the LSP periodograms that resulted from the analysis of the ZTF-{\it g} band (upper panel) and ZTF-{\it r} band (lower panel).  We have detected four signals that exceed the 3.5 $\sigma$ threshold in the ZTF-{\it g} light curve. Their corresponding periods are 119.4 $\pm$ 5.8 d, 188.6 $\pm$ 14.2 d, 224.0 $\pm$ 14.2 d and 325.8 $\pm$ 23.3 d (the values after the plus/minus sign indicate the full width at half maximum (FWHM) of the peaks), respectively. As for the ZTF-{\it r} band, we have detected a single signal that exceeded the 3.5 $\sigma$ threshold, with a period value of 177.9 $\pm$ 12.1 d. We also detect corresponding periods in ZTF-{\it r} band similiar with that in ZTF-{\it g} band, but these signals do not exceeded the 3.5 $\sigma$ threshold.

The longest period is most likely an artifact of the data, as the spectral window function gives a strong peak at 0.0031 c d$^{-1}$ and can be attributed to the annual restriction in the visibility of the source due to its proximity to the Sun. The shorter period is barely significant and is only detected in the ZTF-{\it g} band. It could be a subharmonic of the 225 d signal. The other two significant periods correspond to the quasi-sinusoidal variability that is clearly appreciated in Fig.~\ref{ZTF}. Owing to the irregular sampling of the light curves and the uncertainty involved, these two periodicities probably correspond to the same physical phenomenon. 

Periods longer than the orbital period of the binary have been reported for different type of sources, namely ULX pulsars, disk-fed supergiant X-ray binaries (SGXB), and double-periodic variables \citep{2020MNRAS.495L.139T}, BeXBs \citep{2011MNRAS.413.1600R},  wind-fed SGXBs \citep{2013ApJ...778...45C}, and even black-hole binaries \citep{2012MNRAS.420.1575K}.
\citet{2020MNRAS.495L.139T} suggested that the most likely cause of the observed linear relationship between the orbital period and superorbital periods in high-mass X-ray binary systems is the modulation of precessing hotspots or density waves in accretion disk or circumstellar disk by the binary motion. They reported a distinct correlation between orbital and superorbital periods (see Fig. 1 in \citealt{2020MNRAS.495L.139T}). However, in the case of KS 1947+300, the detected periods do not follow that relationship. For an orbital period of 40 days, a superorbital period of $\sim1000$ days should be observed to agree with the correlation between orbital and superobital periods. Therefore, if the model that attribute the correlation to processes that take place inside the decretion disk is correct, then we conclude that the origin of the periodicities that we detect in the optical light curves of KS 1947+300 must be outside the decretion disk.

An alternative explanation would attribute these periods to some kind of circumbinary motion. Hydrodynamical simulations of the disk evolution in BeXBs showed that most of the material that is lost from the disk is accreted on to the Be star, a small fraction is transferred on to the neutron star and give rise to the X-ray outbursts, and some forms a circumbinary disk around the binary \citep{2019ApJ...881L..32F,2021ApJ...922L..37M}. After several neutron star passages, the decretion disk becomes highly disrupted, with chunks of material more dense than others and/or the formation of a spiral arm. The intensity modulation that we detected could come from the super-orbital motion of a circumbinary material clump or the precession of a spiral arm. A closer look at Fig.~\ref{ZTF} reveals that the largest amplitude of variability is caused by a decrease in brightness (rather than by an increase). There is a pedestal level around $g=14.8$ mag and $r=13.98$ and drops in brightness of $\sim0.15-0.25$ mag. This behavior points toward the occultation of these structures by the Be star.

A third possibility is that it is the disk itself that evolves. It may evolve in physical size, e.g. it expands and partially dissipates on time scales of months or in density. In this third scenario we would expect a corresponding increase and decrease in EW(H$\alpha$). Unfortunately, the spectroscopic observations suffer from numerous gaps. However, Fig.~\ref{ks_multiplot} provides a clue that this may indeed be the case. If we compare the evolution of EW(H$\alpha$) from MJD$\sim$59000 with the ZTF light curves, we observe certain correlation with increases in EW(H$\alpha$) in coincidence with the brightening of the source, e.g. on MJD$~\sim$59100--59200 and MJD$~\sim$59700--59800. This increase in brightness is also seen in the \emph{NEOWISE} near IR data and EW(He I $\lambda$6678). Despite the observational gaps, this quasi-simultaneous changes in the optical and IR continuum and in the line emission of two lines that probe different regions of the decretion disk strongly imply that the origin of the variability is the disk itself.

Our study of KS 1947+300 emphasizes the variable nature of BeXBs in the optical band on all time scales: long (years) as shown in Sect.~\ref{long-term}, intermediate (weeks to months) as shown in this section, and short (hours to days) as shown by \citet{2022A&A...667A..18R}.

\subsection{Color-magnitude diagram} \label{sec:ks_Discu_B-V/V}

\begin{figure}
   \centering
   \includegraphics[width=\hsize]{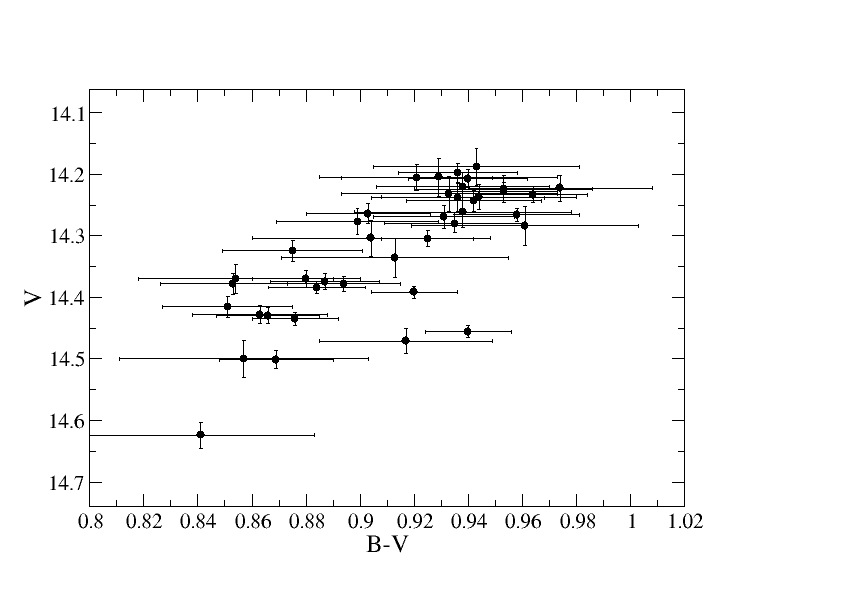}
   \caption{Relationship between $(B-V)$ color index and $V$-band magnitude.}
         \label{ks_B-V_V}
   \end{figure}

\begin{figure}
   \centering
   \includegraphics[width=11cm]{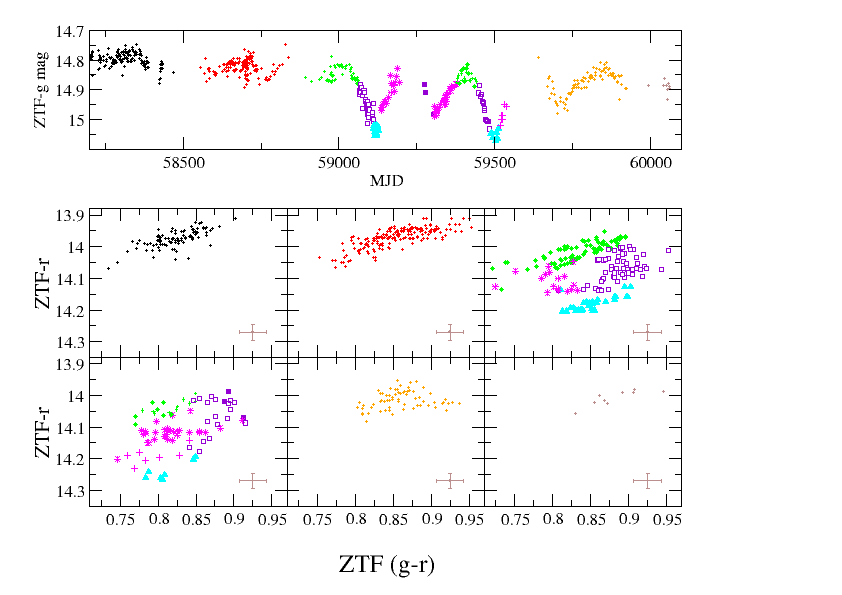}
   \caption{ZTF light curve and color-magnitude diagram. The top panel shows the light curve of the ZTF-{\it g} band. Typical errors are 0.018 mag. The six panels below show the color-magnitude diagram for each individual data set. The color code and symbols in the color-magnitude diagram correspond to those displayed in the light curve.}
        \label{CMD}
   \end{figure}






Figure~\ref{ks_B-V_V} illustrates the relationship between the $(B-V)$ color index and the V-band magnitude.
This type of plot can be used to constrain the inclination angle of the system \citep{1983HvaOB...7...55H}.
A positive correlation is observed in systems where the optical intensity increases (or equivalently, as EW(H$\alpha$) increases) and the radiation becomes redder (i.e., $(B-V)$ increases) with the formation of the circumstellar disk. This correlation is believed to be associated with small or moderate inclination angles \citep{1983HvaOB...7...55H,2015A&A...574A..33R}.

The high cadence of the ZTF data allows us to investigate the evolution of the source in the color-magnitude diagram (CMD) with more detail and may provide a clue about the relatively large scattering observed in Fig.~\ref{ks_B-V_V}. Figure~\ref{CMD} shows the ZTF-{\it g} and color-magnitude diagram $(g-r)-r$. The light curve has been divided into six data sets. The gaps are caused by the lack of data due to the proximity of the source to the Sun. The smaller panels in this figure display the location of the source in the CMD for each date set. As in the case of the broad-band Johnson filters, as the source becomes brighter, it also becomes redder. Interestingly, when the source exhibits large amplitude changes in brightness, it performs a complete loop in the CMD, clockwise. 

If we assume that the brightness variability is due to the evolution of the circumstellar disk, as explained in Sect.~\ref{sec:ks_Discu_periodic_signals_in_ZTF}, then the declining part would correspond to the dissipation of the disk and the rising part to its growth. The CMD reveals that the emission from the system (star plus disk) is redder during the declining phase (violet square symbols in Fig.~\ref{CMD}) and bluer during the rising phase (magenta stars). This result can be understood if the dissipation and subsequent growth starts in the inner parts of the disk. 

The disk contribution to the $B$ and $V$-band emission comes from  very close to the star \citep{2012ApJ...756..156H}. The effective wavelength of the ZTF-{\it g} filter is around 4800 \AA\ (i.e. in between Johnson $B$ and $V$ filters), while that of the ZTF-{\it r} filter is around 6500 \AA\ (i.e. $\sim R$ band) \citep{2019PASP..131a8002B}. If the mechanism that ejects matter from the photosphere of the Be star into the disk ceases (dissipation phase), then a hole will be created in the inner parts of the disk and the emission at longer wavelength will dominate. When this mechanism re-starts, the contribution at shorter wavelength will increase (brightening phase).

\section{Conclusions}\label{sec:ks_Conclusion}

We have studied the evolution of the optical emission of KS 1947+300 since 2000. This long time span allowed us to investigate the X-ray and optical/IR emission in a correlated way. We have shown a long-term fading of the source in the optical/IR band that began after the 2000 X-ray outburst and became more evident after the type II 2013 X-ray outburst. We interpret this decrease of the optical emission as an overall weakening of the decretion disk around the Be star. Surprisingly, in addition to type II X-ray outbursts, KS 1947+300 also displays type I outburst. Because of its nearly circular orbit, KS 1947+300 is not expected to show this kind of periodic outbursts. The fact that we do observe them implies that the disk around the Be star becomes eccentric. As a result, the neutron star is able to capture material when it passes the disk apastron, thus producing type I outbursts. The asymmetric geometric configuration of the disk may explain the fact that type I outbursts in KS 1947+300 do not show a clear preference for a specific orbital phase
(i.e. are not strictly modulated by the orbital period).

We have reported for the first time optical variability on intermediate time scales. KS 1947+300 went through several brightening/fading cycles with characteristic time scales of few months. We studied several scenarios (precession of hot spots or density waves in the decretion disk, circumbinary material clumps), but we concluded that the most likely origin of this variability is changes in the geometrical size of the disk, caused by the temporary halt of the mechanism that ejects material from the photosphere of the Be star into the disk.

Finally, the relationship between the $(B-V)$ color index and the $V$-band magnitude indicated that the inclination angle of KS 1947+300 is small or moderate.

\begin{acknowledgements}

      We acknowledge the support of the staff of the Xinglong
      2.16m/80cm/60cm telescope. This work was partially supported by the Open Project Program of the Key Laboratory of Optical Astronomy, National Astronomical Observatories, Chinese Academy of Sciences.
      
      We acknowledge the support of the staff of the Lijiang 2.4 m telescope. Funding for the telescope has been provided by CAS and the People's Government of Yunnan Province.
      
      Skinakas Observatory is run by the University of Crete and the Foundation for Research and Technology-Hellas.
      
      This research has made use of data provided by the Yaoan High Precision Telescope.

      We acknowledge the Samuel Oschin 48-inch Telescope at the Palomar Observatory as part of the Zwicky Transient Facility project, supported by the National Science Foundation under Grant No. AST-1440341 and a collaboration including Caltech, IPAC, the Weizmann Institute for Science, the Oskar Klein Center at Stockholm University, the University of Maryland, the University of Washington, Deutsches Elektronen-Synchrotron and Humboldt University, Los Alamos National Laboratories, the TANGO Consortium of Taiwan, the University of Wisconsin at Milwaukee, and Lawrence Berkeley National Laboratories. Operations are conducted by COO, IPAC, and UW.

      This publication makes use of data products from \emph{NEOWISE}, which is a project of the Jet Propulsion Laboratory/California Institute of Technology, funded by the Planetary Science Division of the National Aeronautics and Space Administration.

      This research has made use of data and results provided by the ASM/\emph{RXTE} teams at MIT and at the \emph{RXTE} SOF and GOF at NASAs GSFC. Data were obtained through the High Energy Astrophysics Science Archive Research Center Online Service, provided by the NASA/Goddard Space Flight Center.
      This research utilized \emph{MAXI} data provided by RIKEN, JAXA, and the \emph{MAXI} team.
      \emph{Swift}-BAT transient monitor results provided by the \emph{Swift}-BAT team. \emph{Fermi}-GBM results provided by the Fermi Science Support Center.
      
      This work is supported by the National Key R\&D Program of China (2021YFA0718500), and the National Natural Science Foundation of China (Grant No U2031205 and 12233002 ). 
      
      
\end{acknowledgements}

\bibliographystyle{aa} 
\bibliography{KS1947+300.bib} 

\begin{thebibliography}{42}
\expandafter\ifx\csname natexlab\endcsname\relax\def\natexlab#1{#1}\fi

\bibitem[{{Bellm} {et~al.}(2019){Bellm}, {Kulkarni}, {Graham}, {Dekany},
  {Smith}, {Riddle}, {Masci}, {Helou}, {Prince}, {Adams}, {Barbarino},
  {Barlow}, {Bauer}, {Beck}, {Belicki}, {Biswas}, {Blagorodnova}, {Bodewits},
  {Bolin}, {Brinnel}, {Brooke}, {Bue}, {Bulla}, {Burruss}, {Cenko}, {Chang},
  {Connolly}, {Coughlin}, {Cromer}, {Cunningham}, {De}, {Delacroix}, {Desai},
  {Duev}, {Eadie}, {Farnham}, {Feeney}, {Feindt}, {Flynn}, {Franckowiak},
  {Frederick}, {Fremling}, {Gal-Yam}, {Gezari}, {Giomi}, {Goldstein},
  {Golkhou}, {Goobar}, {Groom}, {Hacopians}, {Hale}, {Henning}, {Ho}, {Hover},
  {Howell}, {Hung}, {Huppenkothen}, {Imel}, {Ip}, {Ivezi{\'c}}, {Jackson},
  {Jones}, {Juric}, {Kasliwal}, {Kaspi}, {Kaye}, {Kelley}, {Kowalski},
  {Kramer}, {Kupfer}, {Landry}, {Laher}, {Lee}, {Lin}, {Lin}, {Lunnan},
  {Giomi}, {Mahabal}, {Mao}, {Miller}, {Monkewitz}, {Murphy}, {Ngeow},
  {Nordin}, {Nugent}, {Ofek}, {Patterson}, {Penprase}, {Porter}, {Rauch},
  {Rebbapragada}, {Reiley}, {Rigault}, {Rodriguez}, {van Roestel}, {Rusholme},
  {van Santen}, {Schulze}, {Shupe}, {Singer}, {Soumagnac}, {Stein}, {Surace},
  {Sollerman}, {Szkody}, {Taddia}, {Terek}, {Van Sistine}, {van Velzen},
  {Vestrand}, {Walters}, {Ward}, {Ye}, {Yu}, {Yan}, \&
  {Zolkower}}]{2019PASP..131a8002B}
{Bellm}, E.~C., {Kulkarni}, S.~R., {Graham}, M.~J., {et~al.} 2019, \pasp, 131,
  018002

\bibitem[{{Borozdin} {et~al.}(1990){Borozdin}, {Gilfanov}, {Sunyaev},
  {Churazov}, {Loznikov}, {Yamburenko}, {Skinner}, {Patterson}, {Willmore},
  {Emam}, {Brinkman}, {Heise}, {Int-Zand}, \& {Jager}}]{1990SvAL...16..345B}
{Borozdin}, K., {Gilfanov}, M., {Sunyaev}, R., {et~al.} 1990, Soviet Astronomy
  Letters, 16, 345

\bibitem[{{Chakrabarty} {et~al.}(1995){Chakrabarty}, {Koh}, {Bildsten},
  {Prince}, {Finger}, {Wilson}, {Pendleton}, \& {Rubin}}]{1995ApJ...446..826C}
{Chakrabarty}, D., {Koh}, T., {Bildsten}, L., {et~al.} 1995, \apj, 446, 826

\bibitem[{{Corbet} \& {Krimm}(2013)}]{2013ApJ...778...45C}
{Corbet}, R. H.~D. \& {Krimm}, H.~A. 2013, \apj, 778, 45

\bibitem[{{Doroshenko} {et~al.}(2020){Doroshenko}, {Piraino}, {Doroshenko}, \&
  {Santangelo}}]{2020MNRAS.493.3442D}
{Doroshenko}, R., {Piraino}, S., {Doroshenko}, V., \& {Santangelo}, A. 2020,
  \mnras, 493, 3442

\bibitem[{{Emmanoulopoulos} {et~al.}(2013){Emmanoulopoulos}, {McHardy}, \&
  {Papadakis}}]{2013MNRAS.433..907E}
{Emmanoulopoulos}, D., {McHardy}, I.~M., \& {Papadakis}, I.~E. 2013, \mnras,
  433, 907

\bibitem[{{Fan} {et~al.}(2016){Fan}, {Wang}, {Jiang}, {Wu}, {Li}, {Huang},
  {Xu}, {Hu}, {Zhu}, {Wang}, {Komossa}, \& {Zhang}}]{2016PASP..128k5005F}
{Fan}, Z., {Wang}, H., {Jiang}, X., {et~al.} 2016, \pasp, 128, 115005

\bibitem[{{Franchini} \& {Martin}(2019)}]{2019ApJ...881L..32F}
{Franchini}, A. \& {Martin}, R.~G. 2019, \apjl, 881, L32

\bibitem[{{Fu} {et~al.}(2015){Fu}, {Lubow}, \& {Martin}}]{2015ApJ...807...75F}
{Fu}, W., {Lubow}, S.~H., \& {Martin}, R.~G. 2015, \apj, 807, 75

\bibitem[{{F{\"u}rst} {et~al.}(2014){F{\"u}rst}, {Pottschmidt}, {Wilms},
  {Kennea}, {Bachetti}, {Bellm}, {Boggs}, {Chakrabarty}, {Christensen},
  {Craig}, {Hailey}, {Harrison}, {Stern}, {Tomsick}, {Walton}, \&
  {Zhang}}]{2014ApJ...784L..40F}
{F{\"u}rst}, F., {Pottschmidt}, K., {Wilms}, J., {et~al.} 2014, \apjl, 784, L40

\bibitem[{{Galloway} {et~al.}(2004){Galloway}, {Morgan}, \&
  {Levine}}]{2004ApJ...613.1164G}
{Galloway}, D.~K., {Morgan}, E.~H., \& {Levine}, A.~M. 2004, \apj, 613, 1164

\bibitem[{{Goranskii} {et~al.}(1991){Goranskii}, {Esipov}, {Lyutyi}, \&
  {Shugarov}}]{1991SvAL...17..399G}
{Goranskii}, V.~P., {Esipov}, V.~F., {Lyutyi}, V.~M., \& {Shugarov}, S.~Y.
  1991, Soviet Astronomy Letters, 17, 399

\bibitem[{{Grankin} {et~al.}(1991){Grankin}, {Shevchenko}, \&
  {Yakubov}}]{1991SvAL...17..415G}
{Grankin}, K.~N., {Shevchenko}, V.~S., \& {Yakubov}, S.~D. 1991, Soviet
  Astronomy Letters, 17, 415

\bibitem[{{Harmanec}(1983)}]{1983HvaOB...7...55H}
{Harmanec}, P. 1983, Hvar Observatory Bulletin, 7, 55

\bibitem[{{Haubois} {et~al.}(2012){Haubois}, {Carciofi}, {Rivinius}, {Okazaki},
  \& {Bjorkman}}]{2012ApJ...756..156H}
{Haubois}, X., {Carciofi}, A.~C., {Rivinius}, T., {Okazaki}, A.~T., \&
  {Bjorkman}, J.~E. 2012, \apj, 756, 156

\bibitem[{{Huang} {et~al.}(2012){Huang}, {Li}, {Wang}, {Shang}, {Zhang}, {Hu},
  {Qiu}, \& {Jiang}}]{2012RAA....12.1585H}
{Huang}, F., {Li}, J.-Z., {Wang}, X.-F., {et~al.} 2012, Research in Astronomy
  and Astrophysics, 12, 1585

\bibitem[{{K{\i}z{\i}lo{\u{g}}lu} {et~al.}(2007){K{\i}z{\i}lo{\u{g}}lu},
  {Baykal}, \& {K{\i}z{\i}lo{\u{g}}lu}}]{2007AN....328..142K}
{K{\i}z{\i}lo{\u{g}}lu}, {\"U}., {Baykal}, A., \& {K{\i}z{\i}lo{\u{g}}lu}, N.
  2007, Astronomische Nachrichten, 328, 142

\bibitem[{{Kotze} \& {Charles}(2012)}]{2012MNRAS.420.1575K}
{Kotze}, M.~M. \& {Charles}, P.~A. 2012, \mnras, 420, 1575

\bibitem[{{Krimm} {et~al.}(2013){Krimm}, {Holland}, {Corbet}, {Pearlman},
  {Romano}, {Kennea}, {Bloom}, {Barthelmy}, {Baumgartner}, {Cummings},
  {Gehrels}, {Lien}, {Markwardt}, {Palmer}, {Sakamoto}, {Stamatikos}, \&
  {Ukwatta}}]{2013ApJS..209...14K}
{Krimm}, H.~A., {Holland}, S.~T., {Corbet}, R.~H.~D., {et~al.} 2013, \apjs,
  209, 14

\bibitem[{{Landolt}(2009)}]{2009AJ....137.4186L}
{Landolt}, A.~U. 2009, \aj, 137, 4186

\bibitem[{{Levine} \& {Corbet}(2000)}]{2000IAUC.7523....2L}
{Levine}, A. \& {Corbet}, R. 2000, \iaucirc, 7523, 2

\bibitem[{{Lomb}(1976)}]{1976Ap&SS..39..447L}
{Lomb}, N.~R. 1976, \apss, 39, 447

\bibitem[{{Mainzer} {et~al.}(2011){Mainzer}, {Bauer}, {Grav}, {Masiero},
  {Cutri}, {Dailey}, {Eisenhardt}, {McMillan}, {Wright}, {Walker}, {Jedicke},
  {Spahr}, {Tholen}, {Alles}, {Beck}, {Brandenburg}, {Conrow}, {Evans},
  {Fowler}, {Jarrett}, {Marsh}, {Masci}, {McCallon}, {Wheelock}, {Wittman},
  {Wyatt}, {DeBaun}, {Elliott}, {Elsbury}, {Gautier}, {Gomillion}, {Leisawitz},
  {Maleszewski}, {Micheli}, \& {Wilkins}}]{2011ApJ...731...53M}
{Mainzer}, A., {Bauer}, J., {Grav}, T., {et~al.} 2011, \apj, 731, 53

\bibitem[{{Martin} \& {Franchini}(2021)}]{2021ApJ...922L..37M}
{Martin}, R.~G. \& {Franchini}, A. 2021, \apjl, 922, L37

\bibitem[{{Martin} {et~al.}(2014){Martin}, {Nixon}, {Lubow}, {Armitage},
  {Price}, {Do{\u{g}}an}, \& {King}}]{2014ApJ...792L..33M}
{Martin}, R.~G., {Nixon}, C., {Lubow}, S.~H., {et~al.} 2014, \apjl, 792, L33

\bibitem[{{Meegan} {et~al.}(2009){Meegan}, {Lichti}, {Bhat}, {Bissaldi},
  {Briggs}, {Connaughton}, {Diehl}, {Fishman}, {Greiner}, {Hoover}, {van der
  Horst}, {von Kienlin}, {Kippen}, {Kouveliotou}, {McBreen}, {Paciesas},
  {Preece}, {Steinle}, {Wallace}, {Wilson}, \&
  {Wilson-Hodge}}]{2009ApJ...702..791M}
{Meegan}, C., {Lichti}, G., {Bhat}, P.~N., {et~al.} 2009, \apj, 702, 791

\bibitem[{{Negueruela} {et~al.}(2003){Negueruela}, {Israel}, {Marco}, {Norton},
  \& {Speziali}}]{2003A&A...397..739N}
{Negueruela}, I., {Israel}, G.~L., {Marco}, A., {Norton}, A.~J., \& {Speziali},
  R. 2003, \aap, 397, 739

\bibitem[{{Negueruela} \& {Okazaki}(2001)}]{2001A&A...369..108N}
{Negueruela}, I. \& {Okazaki}, A.~T. 2001, \aap, 369, 108

\bibitem[{{Negueruela} {et~al.}(2001){Negueruela}, {Okazaki}, {Fabregat},
  {Coe}, {Munari}, \& {Tomov}}]{2001A&A...369..117N}
{Negueruela}, I., {Okazaki}, A.~T., {Fabregat}, J., {et~al.} 2001, \aap, 369,
  117

\bibitem[{{Okazaki} \& {Negueruela}(2001)}]{2001A&A...377..161O}
{Okazaki}, A.~T. \& {Negueruela}, I. 2001, \aap, 377, 161

\bibitem[{{Rajoelimanana} {et~al.}(2011){Rajoelimanana}, {Charles}, \&
  {Udalski}}]{2011MNRAS.413.1600R}
{Rajoelimanana}, A.~F., {Charles}, P.~A., \& {Udalski}, A. 2011, \mnras, 413,
  1600

\bibitem[{{Reig}(2008)}]{2008A&A...489..725R}
{Reig}, P. 2008, \aap, 489, 725

\bibitem[{{Reig}(2011)}]{2011Ap&SS.332....1R}
{Reig}, P. 2011, \apss, 332, 1

\bibitem[{{Reig} \& {Fabregat}(2015)}]{2015A&A...574A..33R}
{Reig}, P. \& {Fabregat}, J. 2015, \aap, 574, A33

\bibitem[{{Reig} \& {Fabregat}(2022)}]{2022A&A...667A..18R}
{Reig}, P. \& {Fabregat}, J. 2022, \aap, 667, A18

\bibitem[{{Reig} {et~al.}(2016){Reig}, {Nersesian}, {Zezas}, {Gkouvelis}, \&
  {Coe}}]{2016A&A...590A.122R}
{Reig}, P., {Nersesian}, A., {Zezas}, A., {Gkouvelis}, L., \& {Coe}, M.~J.
  2016, \aap, 590, A122

\bibitem[{{Scargle}(1982)}]{1982ApJ...263..835S}
{Scargle}, J.~D. 1982, \apj, 263, 835

\bibitem[{{Swank} \& {Morgan}(2000)}]{2000IAUC.7531....4S}
{Swank}, J. \& {Morgan}, E. 2000, \iaucirc, 7531, 4

\bibitem[{{Telting} {et~al.}(1994){Telting}, {Heemskerk}, {Henrichs}, \&
  {Savonije}}]{1994A&A...288..558T}
{Telting}, J.~H., {Heemskerk}, M.~H.~M., {Henrichs}, H.~F., \& {Savonije},
  G.~J. 1994, \aap, 288, 558

\bibitem[{{Townsend} \& {Charles}(2020)}]{2020MNRAS.495L.139T}
{Townsend}, L.~J. \& {Charles}, P.~A. 2020, \mnras, 495, 139

\bibitem[{{Tsygankov} \& {Lutovinov}(2005)}]{2005AstL...31...88T}
{Tsygankov}, S.~S. \& {Lutovinov}, A.~A. 2005, Astronomy Letters, 31, 88

\bibitem[{{Zechmeister} \& {K{\"u}rster}(2009)}]{2009A&A...496..577Z}
{Zechmeister}, M. \& {K{\"u}rster}, M. 2009, \aap, 496, 577

\end{thebibliography}

\begin{appendix}


\section{Tables of spectroscopic observations}

\begin{table*}
\caption{Spectroscopic observations of KS 1947+300 from 2.16 m and 2.4 m telescopes.}
\label{table_ks_spec_2m}
 \centering
\begin{tabular}{cclccc}
\hline\hline
Date & MJD & Telescope/ & Wavelength Range & EW(H$\alpha$) & EW(He I $\lambda$6678) \\
(DD-MM-YYYY) & & Instrument & (\AA) & (\AA) & (\AA) \\
\hline
 04-10-2013 & 56569.5298 & 2.16 m/BFOSC & 3000--8600 & -13.5 $\pm$ 0.1 & -0.27 $\pm$ 0.05 \\
 26-10-2013 & 56591.4709 & 2.16 m/OMR   & 5500--6900 & -13.0 $\pm$ 0.2 & -0.24 $\pm$ 0.05 \\
 29-10-2013 & 56594.4809 & 2.16 m/OMR   & 5500--6900 & -13.9 $\pm$ 0.1 & -0.24 $\pm$ 0.30 \\
 17-11-2013 & 56613.4826 & 2.4 m/YFOSC  & 4970--9830 & -13.4 $\pm$ 0.2 & -0.19 $\pm$ 0.12 \\
 18-11-2013 & 56614.5221 & 2.4 m/YFOSC  & 4970--9830 & -14.0 $\pm$ 0.1 & -0.30 $\pm$ 0.15 \\
 19-11-2013 & 56615.5067 & 2.4 m/YFOSC  & 4970--9830 & -14.2 $\pm$ 0.2 & -0.24 $\pm$ 0.13 \\
 20-11-2013 & 56616.5489 & 2.4 m/YFOSC  & 4970--9830 & -14.2 $\pm$ 0.1 & -0.30 $\pm$ 0.14 \\
 21-11-2013 & 56617.5169 & 2.4 m/YFOSC  & 4970--9830 & -15.1 $\pm$ 0.1 & -0.33 $\pm$ 0.14 \\
 22-11-2013 & 56618.5163 & 2.4 m/YFOSC  & 4970--9830 & -14.5 $\pm$ 0.1 & -0.37 $\pm$ 0.15 \\
 23-11-2013 & 56619.5049 & 2.4 m/YFOSC  & 4970--9830 & -14.3 $\pm$ 0.1 & -0.30 $\pm$ 0.16 \\
 24-11-2013 & 56620.5087 & 2.4 m/YFOSC  & 4970--9830 & -14.0 $\pm$ 0.1 & -0.25 $\pm$ 0.13 \\
 12-05-2014 & 56789.8602 & 2.4 m/YFOSC  & 4970--9830 & -14.2 $\pm$ 0.2 & -0.34 $\pm$ 0.15 \\
 20-05-2014 & 56797.7249 & 2.16 m/BFOSC & 5800--8280 & -15.2 $\pm$ 0.1 & -0.39 $\pm$ 0.13 \\
 21-05-2014 & 56798.7677 & 2.16 m/BFOSC & 5800--8280 & -15.4 $\pm$ 0.2 & -0.24 $\pm$ 0.11 \\
 24-06-2014 & 56832.6864 & 2.4 m/YFOSC  & 4970--9830 & -14.1 $\pm$ 0.5 & -0.29 $\pm$ 0.15 \\
 26-06-2014 & 56834.6520 & 2.4 m/YFOSC  & 4970--9830 & -11.2 $\pm$ 0.1 & -0.27 $\pm$ 0.09 \\
 17-09-2014 & 56917.5525 & 2.16 m/OMR   & 5500--6900 & -14.5 $\pm$ 0.1 & -0.23 $\pm$ 0.13 \\
 29-09-2014 & 56929.5746 & 2.16 m/OMR   & 5500--6900 & -13.3 $\pm$ 0.3 & -0.52 $\pm$ 0.22 \\
 12-07-2015 & 57215.5852 & 2.16 m/BFOSC & 5800--8280 & -10.5 $\pm$ 0.2 & -0.11 $\pm$ 0.05 \\
 01-10-2015 & 57296.5719 & 2.16 m/BFOSC & 5800--8280 & -11.7 $\pm$ 0.1 & -0.12 $\pm$ 0.07 \\
 03-10-2015 & 57298.5649 & 2.16 m/BFOSC & 4000--6670 & -12.0 $\pm$ 0.4 & -0.15 $\pm$ 0.08 \\
 04-10-2015 & 57299.5482 & 2.16 m/BFOSC & 5800--8280 & -12.1 $\pm$ 0.1 & -0.17 $\pm$ 0.09 \\
 05-10-2015 & 57300.5707 & 2.16 m/BFOSC & 5800--8280 & -12.4 $\pm$ 0.1 & -0.11 $\pm$ 0.08 \\
 06-10-2015 & 57301.5581 & 2.16 m/BFOSC & 5800--8280 & -12.9 $\pm$ 0.1 & -0.21 $\pm$ 0.16 \\
 08-10-2015 & 57303.5513 & 2.16 m/BFOSC & 5800--8280 & -13.5 $\pm$ 0.1 & -0.32 $\pm$ 0.10 \\
 13-11-2015 & 57339.5946 & 2.4 m/YFOSC  & 4970--9830 & -11.8 $\pm$ 0.2 & -0.14 $\pm$ 0.07 \\
 14-11-2015 & 57340.5625 & 2.4 m/YFOSC  & 4970--9830 & -12.8 $\pm$ 0.1 & -0.16 $\pm$ 0.08 \\
 03-10-2016 & 57664.5125 & 2.16 m/OMR   & 5500--6900 & -12.1 $\pm$ 0.1 & -0.16 $\pm$ 0.49 \\
 09-10-2016 & 57670.5324 & 2.16 m/OMR   & 5500--6900 & -12.3 $\pm$ 0.3 & -0.43 $\pm$ 0.29 \\
 29-10-2017 & 58055.4855 & 2.16 m/OMR   & 5500--6900 & -12.0 $\pm$ 0.1 & -0.43 $\pm$ 0.20 \\
 12-11-2017 & 58069.4499 & 2.16 m/OMR   & 5500--6900 & -12.8 $\pm$ 0.1 & -0.30 $\pm$ 0.22 \\
 23-11-2017 & 58080.5023 & 2.4 m/YFOSC  & 4970--9830 & -13.7 $\pm$ 0.1 & -0.26 $\pm$ 0.10 \\
 25-11-2017 & 58082.4982 & 2.4 m/YFOSC  & 4970--9830 & -13.7 $\pm$ 0.2 & -0.26 $\pm$ 0.10 \\
 15-09-2018 & 58376.5708 & 2.16 m/OMR   & 5500--6900 & -11.8 $\pm$ 0.4 & -0.33 $\pm$ 0.15 \\
 16-09-2018 & 58377.5410 & 2.16 m/OMR   & 5500--6900 & -10.5 $\pm$ 0.5 & -0.26 $\pm$ 0.41 \\
 18-09-2018 & 58379.5969 & 2.16 m/OMR   & 5500--6900 & -10.2 $\pm$ 0.1 & -0.25 $\pm$ 0.25 \\
 20-09-2018 & 58381.5586 & 2.16 m/OMR   & 5500--6900 &  -9.0 $\pm$ 0.9 & -0.19 $\pm$ 0.56 \\
 03-11-2019 & 58790.5159 & 2.16 m/OMR   & 5500--6900 &  -9.5 $\pm$ 0.5 & -0.35 $\pm$ 0.26 \\
 05-11-2019 & 58792.5174 & 2.16 m/OMR   & 5500--6900 &  -9.9 $\pm$ 0.5 & -0.23 $\pm$ 0.46 \\
 09-10-2020 & 59131.5513 & 2.16 m/BFOSC & 5800--8280 & -14.6 $\pm$ 0.2 & -0.71 $\pm$ 0.19 \\
 10-10-2020 & 59132.6247 & 2.16 m/BFOSC & 5800--8280 & -10.2 $\pm$ 0.8 & -0.64 $\pm$ 0.09 \\
 11-10-2020 & 59133.5878 & 2.16 m/BFOSC & 5800--8280 & -10.9 $\pm$ 0.1 & -0.50 $\pm$ 0.13 \\
 12-10-2020 & 59134.5923 & 2.16 m/BFOSC & 5800--8280 & -13.5 $\pm$ 0.3 & -0.94 $\pm$ 0.14 \\
 13-10-2020 & 59135.4988 & 2.16 m/BFOSC & 5800--8280 & -13.5 $\pm$ 0.2 & -0.69 $\pm$ 0.09 \\
\hline
\end{tabular}
\end{table*}

\begin{table*}
\caption{Spectroscopic observations of KS 1947+300 from the 1.3 m telescope (Skinakas observatory).}
\label{table_ks_spec_1.3m_1}
 \centering
\begin{tabular}{cccccc}
\hline\hline
Date & MJD & Telescope & Wavelength Range & EW(H$\alpha$) & EW(He I $\lambda$6678) \\
(DD-MM-YYYY) & & & (\AA) & (\AA) & (\AA) \\
\hline
 30-05-2001 & 52059 & 1.3 m & 5400--7300 & -14.53 $\pm$ 0.73 & -0.19 $\pm$ 0.13 \\
 01-06-2001 & 52061 & 1.3 m & 5400--7300 & -15.10 $\pm$ 0.77 & -0.17 $\pm$ 0.07 \\
 07-08-2001 & 52128 & 1.3 m & 5400--7300 & -14.46 $\pm$ 0.97 & -0.20 $\pm$ 0.05 \\
 09-08-2001 & 52130 & 1.3 m & 5400--7300 & -16.05 $\pm$ 0.70 & -0.25 $\pm$ 0.04 \\
 12-09-2001 & 52164 & 1.3 m & 5400--7300 & -14.38 $\pm$ 0.66 & -0.11 $\pm$ 0.06 \\
 08-10-2001 & 52190 & 1.3 m & 5400--7300 & -15.97 $\pm$ 0.84 & -0.24 $\pm$ 0.08 \\
 17-07-2002 & 52472 & 1.3 m & 5400--7300 & -16.47 $\pm$ 0.85 & -0.07 $\pm$ 0.10 \\
 11-09-2002 & 52528 & 1.3 m & 5400--7300 & -16.04 $\pm$ 0.49 & ... \\
 05-06-2003 & 52795 & 1.3 m & 5400--7300 & -18.11 $\pm$ 0.62 & ... \\
 09-06-2003 & 52799 & 1.3 m & 5400--7300 & -15.79 $\pm$ 0.70 & ... \\
 08-10-2003 & 52920 & 1.3 m & 5400--7300 & -15.60 $\pm$ 0.71 & ... \\
 23-05-2004 & 53148 & 1.3 m & 5400--7300 & -15.69 $\pm$ 0.56 & ... \\
 25-06-2004 & 53181 & 1.3 m & 5400--7300 & -13.51 $\pm$ 0.74 & ... \\
 07-07-2004 & 53193 & 1.3 m & 5400--7300 & -15.41 $\pm$ 0.36 & ... \\
 25-08-2004 & 53242 & 1.3 m & 5400--7300 & -16.06 $\pm$ 0.46 & -0.17 $\pm$ 0.04 \\
 27-08-2004 & 53244 & 1.3 m & 5400--7300 & -16.75 $\pm$ 0.47 & -0.13 $\pm$ 0.05 \\
 03-09-2004 & 53251 & 1.3 m & 5400--7300 & -17.49 $\pm$ 0.56 & -0.47 $\pm$ 0.09 \\
 13-09-2004 & 53261 & 1.3 m & 5400--7300 & -11.85 $\pm$ 0.05 & -0.97 $\pm$ 0.20 \\
 25-10-2004 & 53303 & 1.3 m & 5400--7300 & -14.45 $\pm$ 1.50 & ... \\
 24-05-2005 & 53514 & 1.3 m & 5400--7300 & -16.56 $\pm$ 0.94 & ... \\
 22-06-2005 & 53543 & 1.3 m & 5400--7300 & -15.56 $\pm$ 0.42 & -0.16 $\pm$ 0.05 \\
 11-07-2005 & 53562 & 1.3 m & 5400--7300 & -14.75 $\pm$ 1.49 & ... \\
 16-08-2005 & 53598 & 1.3 m & 5400--7300 & -15.35 $\pm$ 0.51 & -0.12 $\pm$ 0.08 \\
 20-09-2005 & 53633 & 1.3 m & 5400--7300 & -15.17 $\pm$ 0.76 & ... \\
 26-10-2005 & 53669 & 1.3 m & 5400--7300 & -15.30 $\pm$ 0.52 & -0.09 $\pm$ 0.03 \\
 21-06-2006 & 53907 & 1.3 m & 5400--7300 & -14.38 $\pm$ 0.87 & -0.16 $\pm$ 0.06 \\
 15-05-2007 & 54235 & 1.3 m & 5400--7300 & -14.96 $\pm$ 0.97 & -0.10 $\pm$ 0.07 \\
 04-09-2007 & 54347 & 1.3 m & 5400--7300 & -15.83 $\pm$ 0.13 & ... \\
 07-09-2007 & 54350 & 1.3 m & 5400--7300 & -14.25 $\pm$ 0.50 & ... \\
 24-06-2008 & 54641 & 1.3 m & 5400--7300 & -14.73 $\pm$ 0.36 & -0.26 $\pm$ 0.02 \\
 14-07-2008 & 54661 & 1.3 m & 5400--7300 & -14.70 $\pm$ 0.68 & -0.10 $\pm$ 0.04 \\
 08-08-2008 & 54686 & 1.3 m & 5400--7300 & -10.34 $\pm$ 1.08 & ... \\
 03-09-2008 & 54712 & 1.3 m & 5400--7300 & -15.93 $\pm$ 0.50 & ... \\
 08-05-2009 & 54959 & 1.3 m & 5400--7300 & -12.81 $\pm$ 1.40 & ... \\
 30-07-2009 & 55042 & 1.3 m & 5400--7300 & -14.77 $\pm$ 0.51 & -0.06 $\pm$ 0.03 \\
 10-08-2009 & 55053 & 1.3 m & 5400--7300 & -13.26 $\pm$ 2.01 & ... \\
 02-08-2010 & 55410 & 1.3 m & 5400--7300 & -15.06 $\pm$ 0.90 & -0.15 $\pm$ 0.04 \\
 28-08-2010 & 55436 & 1.3 m & 5400--7300 & -14.93 $\pm$ 0.55 & -0.19 $\pm$ 0.01 \\
 04-06-2011 & 55716 & 1.3 m & 5400--7300 & -15.38 $\pm$ 0.63 & -0.14 $\pm$ 0.05 \\
 03-08-2011 & 55776 & 1.3 m & 5400--7300 & -14.46 $\pm$ 0.67 & -0.17 $\pm$ 0.08 \\
 06-06-2012 & 56084 & 1.3 m & 5400--7300 & -14.75 $\pm$ 0.56 & -0.22 $\pm$ 0.06 \\
 24-08-2012 & 56163 & 1.3 m & 5400--7300 & -12.25 $\pm$ 0.53 & ... \\
 31-07-2013 & 56504 & 1.3 m & 5400--7300 & -13.27 $\pm$ 0.67 & -0.13 $\pm$ 0.04 \\
 31-08-2013 & 56535 & 1.3 m & 5400--7300 & -14.86 $\pm$ 0.50 & -0.14 $\pm$ 0.08 \\
 18-10-2013 & 56583 & 1.3 m & 5400--7300 & -12.82 $\pm$ 0.66 & -0.11 $\pm$ 0.02 \\
 06-06-2014 & 56814 & 1.3 m & 5400--7300 & -12.94 $\pm$ 0.74 &  0.04 $\pm$ 0.21 \\
 07-08-2014 & 56876 & 1.3 m & 5400--7300 & -15.97 $\pm$ 0.81 & -0.31 $\pm$ 0.10 \\
 12-10-2014 & 56942 & 1.3 m & 5400--7300 & -13.62 $\pm$ 0.53 & -0.17 $\pm$ 0.04 \\
 23-06-2015 & 57196 & 1.3 m & 5400--7300 & -13.95 $\pm$ 0.60 & -0.18 $\pm$ 0.07 \\
 06-10-2015 & 57301 & 1.3 m & 5400--7300 & -12.43 $\pm$ 0.94 & -0.20 $\pm$ 0.03 \\
 08-06-2016 & 57547 & 1.3 m & 5400--7300 & -11.93 $\pm$ 1.47 & ... \\
 07-09-2016 & 57638 & 1.3 m & 5400--7300 & -14.33 $\pm$ 0.83 & -0.27 $\pm$ 0.21 \\
 26-06-2017 & 57930 & 1.3 m & 5400--7300 & -14.32 $\pm$ 0.89 & -0.39 $\pm$ 0.12 \\
 13-07-2017 & 57947 & 1.3 m & 5400--7300 & -13.35 $\pm$ 1.06 & -0.48 $\pm$ 0.07 \\
 29-08-2017 & 57994 & 1.3 m & 5400--7300 & -14.51 $\pm$ 0.43 & -0.20 $\pm$ 0.09 \\
 15-07-2018 & 58314 & 1.3 m & 5400--7300 & -10.66 $\pm$ 0.83 & -0.03 $\pm$ 0.13 \\
 23-08-2018 & 58353 & 1.3 m & 5400--7300 & -8.55  $\pm$ 0.83 & ... \\
 25-08-2018 & 58355 & 1.3 m & 5400--7300 & -10.14 $\pm$ 0.70 & -0.34 $\pm$ 0.05 \\
\hline
\end{tabular}
\end{table*}

\begin{table*}
\caption{Spectroscopic observations of KS 1947+300 from the 1.3 m telescope (Skinakas observatory)--continued.}
\label{table_ks_spec_1.3m_2}
 \centering
\begin{tabular}{cccccc}
\hline\hline
Date & MJD & Telescope & Wavelength Range & EW(H$\alpha$) & EW(He I $\lambda$6678) \\
(DD-MM-YYYY) & & & (\AA) & (\AA) & (\AA) \\
\hline
 18-09-2018 & 58379 & 1.3 m & 5400--7300 & -9.16  $\pm$ 0.60 & ... \\
 29-07-2019 & 58693 & 1.3 m & 5400--7300 & -7.73  $\pm$ 0.96 & -0.01 $\pm$ 0.07 \\
 09-09-2019 & 58735 & 1.3 m & 5400--7300 & -9.05  $\pm$ 1.02 & -0.21 $\pm$ 0.01 \\
 22-06-2020 & 59022 & 1.3 m & 5400--7300 & -7.77  $\pm$ 0.48 & -2.03 $\pm$ 0.02 \\
 20-07-2020 & 59050 & 1.3 m & 5400--7300 & -6.84  $\pm$ 0.95 & ... \\
 25-08-2020 & 59086 & 1.3 m & 5400--7300 & -8.58  $\pm$ 0.79 & ... \\
 14-09-2020 & 59106 & 1.3 m & 5400--7300 & -12.42 $\pm$ 0.76 & -0.46 $\pm$ 0.08 \\
 16-09-2020 & 59108 & 1.3 m & 5400--7300 & -12.11 $\pm$ 0.84 & -0.57 $\pm$ 0.03 \\
 29-09-2020 & 59121 & 1.3 m & 5400--7300 & -12.94 $\pm$ 1.25 & -0.71 $\pm$ 0.08 \\
 02-07-2021 & 59397 & 1.3 m & 5400--7300 & -8.40  $\pm$ 0.59 & ... \\
 10-08-2021 & 59436 & 1.3 m & 5400--7300 & -7.36  $\pm$ 0.89 & -0.14 $\pm$ 0.02 \\
 01-09-2021 & 59458 & 1.3 m & 5400--7300 & -8.60  $\pm$ 0.68 & -0.28 $\pm$ 0.04 \\
 05-09-2021 & 59462 & 1.3 m & 5400--7300 & -7.74  $\pm$ 1.01 & ... \\
 15-06-2022 & 59745 & 1.3 m & 5400--7300 & -9.83  $\pm$ 0.98 & -0.30 $\pm$ 0.04 \\
 04-07-2022 & 59764 & 1.3 m & 5400--7300 & -10.51 $\pm$ 0.58 & -0.55 $\pm$ 0.06 \\
 30-07-2022 & 59790 & 1.3 m & 5400--7300 & -9.68  $\pm$ 1.13 & -0.31 $\pm$ 0.02 \\
 08-08-2022 & 59799 & 1.3 m & 5400--7300 & -10.47 $\pm$ 0.85 & -0.41 $\pm$ 0.04 \\
 16-08-2022 & 59807 & 1.3 m & 5400--7300 & -9.59  $\pm$ 1.00 & -0.18 $\pm$ 0.08 \\
 08-09-2022 & 59830 & 1.3 m & 5400--7300 & -8.89  $\pm$ 0.89 & ... \\
 19-09-2022 & 59841 & 1.3 m & 5400--7300 & -8.44  $\pm$ 1.00 & -0.28 $\pm$ 0.10 \\
\hline
\end{tabular}
\end{table*}

\section{Tables of photometric observations}

\begin{table*}
\caption{Photometric observations of KS 1947+300 from the 1.3 m telescope (Skinakas observatory).}
\label{table_ks_phot_0}
 \centering
\begin{tabular}{ccccccc}
\hline\hline
 Date & MJD & Telescope & \textit{B} & \textit{V} & \textit{R} & \textit{I} \\
 (DD-MM-YYYY) & & & (mag) & (mag) & (mag) & (mag) \\
\hline
 11-07-2001 & 52102.323  & 1.3 m &15.175 $\pm$ 0.073 & 14.245 $\pm$ 0.078 & 13.565 $\pm$ 0.081 & 12.925 $\pm$ 0.077 \\
 07-06-2003 & 52798.393  & 1.3 m &15.134 $\pm$ 0.031 & 14.205 $\pm$ 0.031 & 13.476 $\pm$ 0.027 & ...                \\  
 05-07-2004 & 53192.303  & 1.3 m &15.147 $\pm$ 0.017 & 14.207 $\pm$ 0.014 & 13.520 $\pm$ 0.017 & 12.857 $\pm$ 0.022 \\
 27-07-2004 & 53214.473  & 1.3 m &15.131 $\pm$ 0.024 & 14.188 $\pm$ 0.030 & 13.486 $\pm$ 0.027 & 12.808 $\pm$ 0.041 \\
 14-09-2004 & 53263.322  & 1.3 m &15.134 $\pm$ 0.015 & 14.198 $\pm$ 0.016 & 13.505 $\pm$ 0.014 & 12.812 $\pm$ 0.026 \\
 28-07-2005 & 53580.455  & 1.3 m &15.127 $\pm$ 0.019 & 14.206 $\pm$ 0.021 & 13.518 $\pm$ 0.021 & ...                \\  
 18-08-2006 & 53966.400  & 1.3 m &15.165 $\pm$ 0.027 & 14.232 $\pm$ 0.029 & 13.531 $\pm$ 0.017 & ...                \\  
 16-07-2007 & 54298.392  & 1.3 m &15.245 $\pm$ 0.028 & 14.284 $\pm$ 0.031 & 13.533 $\pm$ 0.025 & 12.924 $\pm$ 0.029 \\
 01-09-2007 & 54345.481  & 1.3 m &15.197 $\pm$ 0.027 & 14.223 $\pm$ 0.021 & 13.558 $\pm$ 0.025 & 12.877 $\pm$ 0.041 \\
 02-09-2007 & 54346.331  & 1.3 m &15.249 $\pm$ 0.028 & 14.336 $\pm$ 0.031 & 13.601 $\pm$ 0.020 & 12.902 $\pm$ 0.021 \\
 06-08-2008 & 54685.528  & 1.3 m &15.200 $\pm$ 0.031 & 14.262 $\pm$ 0.025 & 13.552 $\pm$ 0.029 & 12.870 $\pm$ 0.032 \\
 09-08-2008 & 54688.302  & 1.3 m &15.091 $\pm$ 0.054 & 14.314 $\pm$ 0.066 & 13.545 $\pm$ 0.037 & 12.862 $\pm$ 0.039 \\
 29-06-2009 & 55012.405  & 1.3 m &15.177 $\pm$ 0.025 & 14.224 $\pm$ 0.022 & 13.554 $\pm$ 0.018 & 12.866 $\pm$ 0.018 \\
 26-07-2010 & 55404.362  & 1.3 m &15.181 $\pm$ 0.030 & 14.237 $\pm$ 0.020 & 13.572 $\pm$ 0.025 & 12.904 $\pm$ 0.028 \\
 26-08-2011 & 55800.383  & 1.3 m &15.181 $\pm$ 0.014 & 14.228 $\pm$ 0.014 & 13.564 $\pm$ 0.015 & 12.881 $\pm$ 0.025 \\
 09-09-2011 & 55814.241  & 1.3 m &15.174 $\pm$ 0.022 & 14.238 $\pm$ 0.023 & 13.592 $\pm$ 0.015 & 12.903 $\pm$ 0.032 \\
 06-11-2012 & 56238.259  & 1.3 m &14.945 $\pm$ 0.273 & 14.141 $\pm$ 0.237 & 13.496 $\pm$ 0.235 & 12.755 $\pm$ 0.241 \\
 29-07-2013 & 56503.310  & 1.3 m &15.185 $\pm$ 0.019 & 14.243 $\pm$ 0.017 & 13.574 $\pm$ 0.019 & 12.921 $\pm$ 0.022 \\
 29-08-2013 & 56534.329  & 1.3 m &15.158 $\pm$ 0.023 & 14.220 $\pm$ 0.022 & 13.528 $\pm$ 0.023 & 12.817 $\pm$ 0.039 \\
 20-08-2014 & 56890.268  & 1.3 m &15.176 $\pm$ 0.021 & 14.277 $\pm$ 0.021 & 13.622 $\pm$ 0.020 & 12.957 $\pm$ 0.027 \\
 14-09-2014 & 56915.333  & 1.3 m &15.215 $\pm$ 0.022 & 14.280 $\pm$ 0.014 & 13.609 $\pm$ 0.017 & 12.939 $\pm$ 0.027 \\
 22-07-2015 & 57226.376  & 1.3 m &15.224 $\pm$ 0.020 & 14.266 $\pm$ 0.011 & 13.611 $\pm$ 0.010 & 12.930 $\pm$ 0.012 \\
 18-11-2015 & 57345.193  & 1.3 m &15.197 $\pm$ 0.015 & 14.233 $\pm$ 0.013 & 13.554 $\pm$ 0.015 & 12.881 $\pm$ 0.015 \\
 06-06-2016 & 57546.362  & 1.3 m &15.167 $\pm$ 0.017 & 14.264 $\pm$ 0.016 & 13.565 $\pm$ 0.017 & 12.898 $\pm$ 0.029 \\
 08-09-2016 & 57640.269  & 1.3 m &15.273 $\pm$ 0.090 & 14.331 $\pm$ 0.084 & 13.679 $\pm$ 0.086 & 13.017 $\pm$ 0.093 \\
 06-10-2016 & 57668.212  & 1.3 m &15.230 $\pm$ 0.011 & 14.305 $\pm$ 0.013 & 13.632 $\pm$ 0.012 & 12.944 $\pm$ 0.013 \\
 03-11-2016 & 57696.223  & 1.3 m &15.200 $\pm$ 0.018 & 14.269 $\pm$ 0.019 & 13.604 $\pm$ 0.020 & 12.925 $\pm$ 0.024 \\
 25-06-2017 & 57930.357  & 1.3 m &15.208 $\pm$ 0.033 & 14.304 $\pm$ 0.029 & 13.605 $\pm$ 0.062 & 12.934 $\pm$ 0.072 \\
 28-08-2017 & 57994.355  & 1.3 m &15.199 $\pm$ 0.020 & 14.324 $\pm$ 0.017 & 13.713 $\pm$ 0.018 & 13.062 $\pm$ 0.025 \\
 13-06-2018 & 58283.358  & 1.3 m &15.188 $\pm$ 0.051 & 14.313 $\pm$ 0.049 & 13.682 $\pm$ 0.056 & 13.038 $\pm$ 0.070 \\
 31-07-2018 & 58331.345  & 1.3 m &15.231 $\pm$ 0.021 & 14.378 $\pm$ 0.017 & 13.752 $\pm$ 0.020 & 13.135 $\pm$ 0.033 \\
 13-07-2019 & 58678.328  & 1.3 m &15.224 $\pm$ 0.027 & 14.370 $\pm$ 0.024 & 13.761 $\pm$ 0.022 & 13.142 $\pm$ 0.036 \\
 30-07-2019 & 58695.300  & 1.3 m &15.250 $\pm$ 0.014 & 14.370 $\pm$ 0.014 & 13.741 $\pm$ 0.011 & 13.116 $\pm$ 0.019 \\
 11-09-2019 & 58738.337  & 1.3 m &15.272 $\pm$ 0.017 & 14.378 $\pm$ 0.012 & 13.751 $\pm$ 0.014 & 13.078 $\pm$ 0.022 \\
 24-06-2020 & 59025.375  & 1.3 m &15.268 $\pm$ 0.016 & 14.384 $\pm$ 0.009 & 13.763 $\pm$ 0.010 & 13.151 $\pm$ 0.013 \\
 19-07-2020 & 59050.320  & 1.3 m &15.312 $\pm$ 0.013 & 14.392 $\pm$ 0.010 & 13.764 $\pm$ 0.012 & 13.126 $\pm$ 0.021 \\
 19-08-2020 & 59081.323  & 1.3 m &15.396 $\pm$ 0.012 & 14.456 $\pm$ 0.010 & 13.806 $\pm$ 0.008 & 13.138 $\pm$ 0.013 \\
 04-07-2021 & 59400.359  & 1.3 m &15.266 $\pm$ 0.017 & 14.415 $\pm$ 0.017 & 13.821 $\pm$ 0.018 & 13.224 $\pm$ 0.022 \\
 04-09-2021 & 59462.354  & 1.3 m &15.388 $\pm$ 0.024 & 14.471 $\pm$ 0.021 & 13.816 $\pm$ 0.022 & 13.167 $\pm$ 0.031 \\
 10-10-2021 & 59498.290  & 1.3 m &15.465 $\pm$ 0.036 & 14.624 $\pm$ 0.021 & 14.010 $\pm$ 0.022 & 13.362 $\pm$ 0.026 \\
 29-07-2022 & 59790.379  & 1.3 m &15.311 $\pm$ 0.013 & 14.435 $\pm$ 0.010 & 13.840 $\pm$ 0.015 & 13.247 $\pm$ 0.025 \\
 17-08-2022 & 59809.330  & 1.3 m &15.296 $\pm$ 0.014 & 14.430 $\pm$ 0.013 & 13.821 $\pm$ 0.019 & 13.225 $\pm$ 0.020 \\
 20-09-2022 & 59843.312  & 1.3 m &15.262 $\pm$ 0.015 & 14.375 $\pm$ 0.013 & 13.776 $\pm$ 0.013 & 13.144 $\pm$ 0.019 \\
 29-06-2023 & 60125.366  & 1.3 m &15.357 $\pm$ 0.035 & 14.500 $\pm$ 0.030 & 13.890 $\pm$ 0.028 & 13.257 $\pm$ 0.025 \\
 12-07-2023 & 60138.348  & 1.3 m &15.370 $\pm$ 0.014 & 14.501 $\pm$ 0.015 & 13.892 $\pm$ 0.013 & 13.265 $\pm$ 0.017 \\
 23-08-2023 & 60180.328  & 1.3 m &15.291 $\pm$ 0.021 & 14.428 $\pm$ 0.014 & 13.846 $\pm$ 0.014 & 13.249 $\pm$ 0.020 \\
\hline
\end{tabular}
\end{table*}

\begin{table*}
\caption{Photometric observations of KS 1947+300 from 60 cm , 80 cm, 2.4 m, and YAHPT telescopes.}
\label{table_ks_phot_1}
 \centering
\begin{tabular}{ccccccc}
\hline\hline
 Date & MJD & Telescope & \textit{B} & \textit{V} & \textit{R} & \textit{I} \\
 (DD-MM-YYYY) & & & (mag) & (mag) & (mag) & (mag) \\
\hline
 04-10-2013 & 56569.51 & 80cm & 15.176 $\pm$ 0.028 & 14.261 $\pm$ 0.030 & 13.627 $\pm$ 0.033 & 12.876 $\pm$ 0.033 \\
 15-10-2013 & 56580.45 & 80cm & 15.095 $\pm$ 0.024 & 14.175 $\pm$ 0.023 & 13.554 $\pm$ 0.023 & 12.811 $\pm$ 0.026 \\
 25-10-2013 & 56590.45 & 80cm & 15.105 $\pm$ 0.009 & 14.198 $\pm$ 0.007 & 13.548 $\pm$ 0.006 & 12.865 $\pm$ 0.009 \\
 26-10-2013 & 56591.45 & 80cm & 15.116 $\pm$ 0.009 & 14.203 $\pm$ 0.007 & 13.545 $\pm$ 0.007 & 12.843 $\pm$ 0.009 \\
 27-10-2013 & 56592.49 & 80cm & ...                & ...                & 13.548 $\pm$ 0.020 & 12.842 $\pm$ 0.012 \\
 29-10-2013 & 56594.45 & 80cm & 15.116 $\pm$ 0.009 & 14.190 $\pm$ 0.007 & 13.551 $\pm$ 0.007 & 12.841 $\pm$ 0.009 \\
 30-10-2013 & 56595.45 & 80cm & 15.137 $\pm$ 0.011 & 14.225 $\pm$ 0.008 & 13.578 $\pm$ 0.007 & 12.860 $\pm$ 0.009 \\
 18-11-2013 & 56614.10 & 2.4m & 15.096 $\pm$ 0.007 & 14.184 $\pm$ 0.006 & 13.483 $\pm$ 0.006 & ...                \\
 19-11-2013 & 56615.02 & 2.4m & 15.094 $\pm$ 0.007 & 14.190 $\pm$ 0.006 & 13.494 $\pm$ 0.006 & ...                \\
 20-11-2013 & 56616.18 & 2.4m & 15.100 $\pm$ 0.007 & 14.190 $\pm$ 0.007 & 13.492 $\pm$ 0.006 & 12.786 $\pm$ 0.009 \\
 21-11-2013 & 56617.77 & 2.4m & 15.100 $\pm$ 0.007 & 14.190 $\pm$ 0.006 & 13.504 $\pm$ 0.006 & 12.808 $\pm$ 0.009 \\
 22-11-2013 & 56618.57 & 2.4m & 15.101 $\pm$ 0.007 & 14.190 $\pm$ 0.006 & 13.500 $\pm$ 0.006 & 12.802 $\pm$ 0.009 \\
 23-11-2013 & 56619.10 & 2.4m & 15.091 $\pm$ 0.007 & 14.194 $\pm$ 0.006 & 13.501 $\pm$ 0.006 & 12.791 $\pm$ 0.009 \\
 07-05-2014 & 56784.81 & 80cm & 15.167 $\pm$ 0.020 & 14.258 $\pm$ 0.015 & 13.615 $\pm$ 0.012 & 12.930 $\pm$ 0.013 \\
 13-05-2014 & 56790.09 & 80cm & 15.162 $\pm$ 0.007 & 14.265 $\pm$ 0.006 & 13.614 $\pm$ 0.006 & 12.906 $\pm$ 0.009 \\
 14-05-2014 & 56791.75 & 80cm & 15.172 $\pm$ 0.028 & 14.269 $\pm$ 0.016 & 13.658 $\pm$ 0.016 & 12.940 $\pm$ 0.013 \\
 05-06-2014 & 56813.16 & 2.4m & 15.189 $\pm$ 0.007 & 14.296 $\pm$ 0.007 & 13.617 $\pm$ 0.006 & 12.940 $\pm$ 0.009 \\
 24-06-2014 & 56832.95 & 2.4m & 15.168 $\pm$ 0.011 & 14.301 $\pm$ 0.007 & ...                & ...                \\
 25-06-2014 & 56833.01 & 2.4m & ...                & ...                & 13.632 $\pm$ 0.008 & 12.946 $\pm$ 0.009 \\
 27-06-2014 & 56835.32 & 2.4m & 15.191 $\pm$ 0.007 & 14.306 $\pm$ 0.007 & 13.622 $\pm$ 0.006 & 12.940 $\pm$ 0.009 \\
 27-09-2014 & 56927.49 & 80cm & 15.176 $\pm$ 0.014 & 14.267 $\pm$ 0.009 & 13.660 $\pm$ 0.009 & 12.990 $\pm$ 0.011 \\
 29-09-2014 & 56929.47 & 80cm & 15.225 $\pm$ 0.018 & 14.271 $\pm$ 0.011 & 13.623 $\pm$ 0.011 & 12.937 $\pm$ 0.013 \\
 30-09-2014 & 56930.48 & 80cm & 15.224 $\pm$ 0.028 & 14.275 $\pm$ 0.014 & 13.636 $\pm$ 0.012 & 12.961 $\pm$ 0.013 \\
 20-03-2015 & 57101.88 & 60cm & 14.974 $\pm$ 0.010 & 14.111 $\pm$ 0.012 & 13.637 $\pm$ 0.018 & ...                \\
 22-03-2015 & 57103.80 & 60cm & 15.006 $\pm$ 0.009 & 14.162 $\pm$ 0.009 & 13.667 $\pm$ 0.008 & 13.186 $\pm$ 0.011 \\
 23-03-2015 & 57104.85 & 60cm & 14.942 $\pm$ 0.009 & 14.102 $\pm$ 0.009 & 13.595 $\pm$ 0.008 & 13.130 $\pm$ 0.011 \\
 07-06-2015 & 57180.66 & 80cm & 15.192 $\pm$ 0.009 & 14.262 $\pm$ 0.008 & 13.617 $\pm$ 0.008 & 12.910 $\pm$ 0.011 \\
 01-10-2015 & 57296.50 & 80cm & 15.218 $\pm$ 0.008 & 14.283 $\pm$ 0.007 & 13.630 $\pm$ 0.008 & 12.922 $\pm$ 0.011 \\
 02-10-2015 & 57297.57 & 80cm & 15.023 $\pm$ 0.018 & 14.129 $\pm$ 0.010 & 13.668 $\pm$ 0.010 & 12.952 $\pm$ 0.012 \\
 03-10-2015 & 57298.46 & 80cm & 15.171 $\pm$ 0.011 & 14.252 $\pm$ 0.008 & 13.586 $\pm$ 0.011 & 12.895 $\pm$ 0.013 \\
 04-10-2015 & 57299.46 & 80cm & 15.191 $\pm$ 0.012 & 14.264 $\pm$ 0.009 & 13.599 $\pm$ 0.012 & 12.857 $\pm$ 0.014 \\
 05-10-2015 & 57300.49 & 80cm & 15.169 $\pm$ 0.011 & 14.241 $\pm$ 0.009 & 13.572 $\pm$ 0.009 & 12.859 $\pm$ 0.012 \\
 08-10-2015 & 57303.47 & 80cm & ...                & 14.292 $\pm$ 0.008 & 13.609 $\pm$ 0.008 & 12.911 $\pm$ 0.010 \\
 31-10-2015 & 57326.32 & 80cm & 15.163 $\pm$ 0.009 & 14.247 $\pm$ 0.009 & 13.578 $\pm$ 0.008 & 12.872 $\pm$ 0.012 \\
 01-11-2015 & 57327.12 & 80cm & 15.176 $\pm$ 0.012 & 14.247 $\pm$ 0.011 & 13.567 $\pm$ 0.010 & 12.877 $\pm$ 0.013 \\
 13-11-2015 & 57339.09 & 2.4m & ...                & ...                & 13.583 $\pm$ 0.006 & 12.885 $\pm$ 0.009 \\
 29-11-2015 & 57355.51 & 80cm & 15.170 $\pm$ 0.014 & 14.211 $\pm$ 0.009 & 13.577 $\pm$ 0.010 & 12.872 $\pm$ 0.011 \\
 30-11-2015 & 57356.50 & 80cm & 15.180 $\pm$ 0.045 & 14.225 $\pm$ 0.020 & 13.541 $\pm$ 0.018 & 12.873 $\pm$ 0.018 \\
 19-09-2016 & 57650.52 & 60cm & 15.154 $\pm$ 0.009 & 14.260 $\pm$ 0.008 & 13.601 $\pm$ 0.007 & 12.898 $\pm$ 0.011 \\
 20-09-2016 & 57651.52 & 60cm & 15.085 $\pm$ 0.008 & 14.251 $\pm$ 0.009 & 13.597 $\pm$ 0.008 & 12.903 $\pm$ 0.011 \\
 27-09-2016 & 57658.52 & 60cm & 15.178 $\pm$ 0.009 & 14.285 $\pm$ 0.009 & 13.576 $\pm$ 0.010 & 12.888 $\pm$ 0.013 \\
 28-09-2016 & 57659.54 & 60cm & 15.137 $\pm$ 0.011 & 14.253 $\pm$ 0.008 & 13.581 $\pm$ 0.008 & 12.898 $\pm$ 0.012 \\
 25-10-2016 & 57686.51 & 60cm & 15.177 $\pm$ 0.012 & 14.272 $\pm$ 0.010 & 13.566 $\pm$ 0.010 & 12.903 $\pm$ 0.013 \\
 29-10-2016 & 57690.53 & 60cm & ...                & ...                & 13.598 $\pm$ 0.009 & 12.900 $\pm$ 0.012 \\
 30-10-2016 & 57692.55 & 60cm & 15.190 $\pm$ 0.010 & 14.282 $\pm$ 0.010 & 13.581 $\pm$ 0.009 & 12.908 $\pm$ 0.012 \\
 23-11-2016 & 57715.49 & 60cm & ...                & ...                & 13.635 $\pm$ 0.009 & 12.930 $\pm$ 0.013 \\
 01-12-2016 & 57723.48 & 60cm & 15.157 $\pm$ 0.009 & 14.295 $\pm$ 0.009 & 13.589 $\pm$ 0.010 & 12.916 $\pm$ 0.013 \\
 15-09-2017 & 58011.54 & 80cm & 15.224 $\pm$ 0.009 & 14.341 $\pm$ 0.008 & 13.706 $\pm$ 0.008 & 13.030 $\pm$ 0.011 \\
 17-09-2017 & 58013.57 & 80cm & 15.214 $\pm$ 0.008 & 14.341 $\pm$ 0.008 & 13.696 $\pm$ 0.007 & 13.027 $\pm$ 0.011 \\
 18-09-2017 & 58014.53 & 80cm & 15.233 $\pm$ 0.008 & 14.340 $\pm$ 0.008 & 13.702 $\pm$ 0.008 & 13.042 $\pm$ 0.010 \\
 19-09-2017 & 58015.54 & 80cm & 15.217 $\pm$ 0.010 & 14.355 $\pm$ 0.009 & 13.725 $\pm$ 0.010 & 13.097 $\pm$ 0.013 \\
 20-09-2017 & 58016.52 & 80cm & 15.235 $\pm$ 0.010 & 14.361 $\pm$ 0.009 & 13.725 $\pm$ 0.008 & 13.071 $\pm$ 0.012 \\
 27-10-2017 & 58053.48 & 60cm & 15.217 $\pm$ 0.012 & 14.323 $\pm$ 0.011 & 13.678 $\pm$ 0.011 & 13.006 $\pm$ 0.014 \\
 28-10-2017 & 58054.48 & 60cm & 15.240 $\pm$ 0.015 & 14.385 $\pm$ 0.012 & 13.759 $\pm$ 0.011 & 13.048 $\pm$ 0.015 \\
 29-10-2017 & 58055.48 & 60cm & 15.198 $\pm$ 0.012 & 14.324 $\pm$ 0.011 & 13.684 $\pm$ 0.010 & 13.023 $\pm$ 0.013 \\
 30-10-2017 & 58056.47 & 60cm & 15.197 $\pm$ 0.012 & 14.334 $\pm$ 0.009 & 13.683 $\pm$ 0.010 & 13.025 $\pm$ 0.012 \\
 19-11-2017 & 58076.44 & 60cm & 15.161 $\pm$ 0.011 & 14.307 $\pm$ 0.010 & 13.671 $\pm$ 0.009 & 13.012 $\pm$ 0.015 \\
\hline
\end{tabular}
\end{table*}

\begin{table*}
\caption{Photometric observations of KS 1947+300 from 60 cm , 80 cm, 2.4 m, and YAHPT telescopes--continued.}
\label{table_ks_phot_2}
 \centering
\begin{tabular}{ccccccc}
\hline\hline
 Date & MJD & Telescope & \textit{B} & \textit{V} & \textit{R} & \textit{I} \\
 (DD-MM-YYYY) & & & (mag) & (mag) & (mag) & (mag) \\
\hline
 20-11-2017 & 58077.42 & 60cm & 15.200 $\pm$ 0.009 & 14.324 $\pm$ 0.008 & 13.674 $\pm$ 0.009 & 13.020 $\pm$ 0.013 \\
 21-11-2017 & 58078.46 & 60cm & 15.196 $\pm$ 0.011 & 14.326 $\pm$ 0.010 & 13.682 $\pm$ 0.010 & 13.026 $\pm$ 0.014 \\
 22-11-2017 & 58079.45 & 60cm & 15.203 $\pm$ 0.011 & 14.306 $\pm$ 0.012 & 13.677 $\pm$ 0.009 & 13.071 $\pm$ 0.050 \\
 23-11-2017 & 58080.48 & 60cm & 15.181 $\pm$ 0.011 & 14.317 $\pm$ 0.009 & 13.673 $\pm$ 0.008 & 13.033 $\pm$ 0.011 \\
 08-09-2018 & 58369.59 & 80cm & 15.249 $\pm$ 0.010 & 14.348 $\pm$ 0.008 & 13.722 $\pm$ 0.007 & 13.077 $\pm$ 0.010 \\
 09-09-2018 & 58370.53 & 80cm & 15.251 $\pm$ 0.010 & 14.359 $\pm$ 0.008 & 13.736 $\pm$ 0.007 & 13.109 $\pm$ 0.010 \\
 16-09-2018 & 58377.52 & 80cm & 15.240 $\pm$ 0.008 & 14.337 $\pm$ 0.007 & 13.712 $\pm$ 0.007 & 13.057 $\pm$ 0.010 \\
 08-11-2018 & 58430.47 & 80cm & 15.266 $\pm$ 0.009 & 14.378 $\pm$ 0.008 & 13.767 $\pm$ 0.008 & 13.135 $\pm$ 0.011 \\
 24-09-2019 & 58750.52 & 80cm & 15.318 $\pm$ 0.008 & 14.439 $\pm$ 0.007 & 13.806 $\pm$ 0.007 & 13.173 $\pm$ 0.009 \\
 25-09-2019 & 58751.52 & 80cm & 15.297 $\pm$ 0.008 & 14.405 $\pm$ 0.007 & 13.780 $\pm$ 0.007 & 13.141 $\pm$ 0.009 \\
 04-10-2019 & 58760.49 & 80cm & 15.366 $\pm$ 0.009 & 14.476 $\pm$ 0.009 & 13.858 $\pm$ 0.008 & 13.212 $\pm$ 0.011 \\
 03-11-2019 & 58790.47 & 80cm & 15.274 $\pm$ 0.010 & 14.390 $\pm$ 0.008 & 13.788 $\pm$ 0.007 & 13.161 $\pm$ 0.010 \\
 04-11-2019 & 58791.45 & 80cm & 15.274 $\pm$ 0.009 & 14.401 $\pm$ 0.008 & 13.792 $\pm$ 0.007 & 13.160 $\pm$ 0.010 \\
 15-09-2020 & 59107.59 & 80cm & 15.443 $\pm$ 0.009 & 14.540 $\pm$ 0.008 & 13.885 $\pm$ 0.007 & 13.211 $\pm$ 0.011 \\
 16-09-2020 & 59108.53 & 80cm & 15.450 $\pm$ 0.009 & 14.557 $\pm$ 0.008 & 13.912 $\pm$ 0.007 & 13.255 $\pm$ 0.010 \\
 09-10-2020 & 59131.49 & 80cm & 15.422 $\pm$ 0.014 & 14.642 $\pm$ 0.011 & 13.921 $\pm$ 0.009 & 13.289 $\pm$ 0.011 \\
 12-10-2020 & 59134.49 & 80cm & 15.422 $\pm$ 0.009 & 14.553 $\pm$ 0.008 & 13.938 $\pm$ 0.007 & 13.318 $\pm$ 0.009 \\
 20-11-2020 & 59173.50 & 80cm & 15.341 $\pm$ 0.022 & 14.473 $\pm$ 0.026 & 13.889 $\pm$ 0.028 & 13.207 $\pm$ 0.032 \\
 22-11-2020 & 59175.42 & 80cm & 15.313 $\pm$ 0.012 & 14.449 $\pm$ 0.009 & 13.854 $\pm$ 0.008 & 13.242 $\pm$ 0.011 \\
 21-12-2020 & 59204.44 & 80cm & 15.263 $\pm$ 0.012 & 14.392 $\pm$ 0.010 & 13.812 $\pm$ 0.008 & 13.193 $\pm$ 0.011 \\
 10-10-2021 & 59497.50 & 80cm & 15.485 $\pm$ 0.011 & 14.612 $\pm$ 0.008 & 13.987 $\pm$ 0.008 & 13.361 $\pm$ 0.011 \\
12-10-2022 & 59864.62 & YAHPT & 15.275 $\pm$ 0.009 & 14.375 $\pm$ 0.007 & 13.773 $\pm$ 0.018 & 13.157 $\pm$ 0.024 \\
13-10-2022 & 59865.51 & YAHPT & 15.262 $\pm$ 0.006 & 14.373 $\pm$ 0.006 & 13.754 $\pm$ 0.006 & 13.144 $\pm$ 0.009 \\
18-10-2022 & 59870.65 & YAHPT & 15.372 $\pm$ 0.097 & 14.107 $\pm$ 0.107 & 13.618 $\pm$ 0.120 & ...                \\
20-10-2022 & 59872.50 & YAHPT & 15.264 $\pm$ 0.006 & 14.373 $\pm$ 0.006 & 13.757 $\pm$ 0.006 & 13.151 $\pm$ 0.009 \\
22-10-2022 & 59874.50 & YAHPT & 15.271 $\pm$ 0.006 & 14.376 $\pm$ 0.006 & 13.758 $\pm$ 0.006 & 13.153 $\pm$ 0.009 \\
28-10-2022 & 59880.65 & YAHPT & 15.279 $\pm$ 0.007 & 14.377 $\pm$ 0.006 & 13.738 $\pm$ 0.006 & 13.121 $\pm$ 0.010 \\
29-10-2022 & 59881.64 & YAHPT & 15.268 $\pm$ 0.007 & 14.362 $\pm$ 0.006 & 13.741 $\pm$ 0.006 & 13.133 $\pm$ 0.009 \\
30-10-2022 & 59882.60 & YAHPT & 15.285 $\pm$ 0.007 & 14.366 $\pm$ 0.006 & 13.744 $\pm$ 0.006 & 13.140 $\pm$ 0.009 \\
01-11-2022 & 59884.59 & YAHPT & 15.292 $\pm$ 0.007 & 14.386 $\pm$ 0.006 & 13.752 $\pm$ 0.006 & 13.149 $\pm$ 0.009 \\
02-11-2022 & 59885.63 & YAHPT & 15.275 $\pm$ 0.008 & 14.380 $\pm$ 0.007 & 13.741 $\pm$ 0.007 & 13.121 $\pm$ 0.010 \\
03-11-2022 & 59886.62 & YAHPT & 15.267 $\pm$ 0.008 & 14.379 $\pm$ 0.007 & 13.757 $\pm$ 0.007 & 13.134 $\pm$ 0.010 \\
06-11-2022 & 59889.63 & YAHPT & 15.268 $\pm$ 0.011 & 14.358 $\pm$ 0.008 & 13.747 $\pm$ 0.007 & 13.101 $\pm$ 0.011 \\
07-11-2022 & 59890.62 & YAHPT & 15.287 $\pm$ 0.016 & 14.352 $\pm$ 0.009 & 13.715 $\pm$ 0.009 & 13.087 $\pm$ 0.013 \\
08-11-2022 & 59891.62 & YAHPT & 15.271 $\pm$ 0.017 & 14.394 $\pm$ 0.009 & 13.745 $\pm$ 0.009 & 13.125 $\pm$ 0.013 \\
09-11-2022 & 59892.61 & YAHPT & 15.273 $\pm$ 0.016 & 14.358 $\pm$ 0.008 & 13.727 $\pm$ 0.007 & 13.120 $\pm$ 0.011 \\
10-11-2022 & 59893.62 & YAHPT & 15.277 $\pm$ 0.012 & 14.368 $\pm$ 0.008 & 13.733 $\pm$ 0.008 & 13.069 $\pm$ 0.012 \\
12-11-2022 & 59895.61 & YAHPT & 15.285 $\pm$ 0.010 & 14.374 $\pm$ 0.008 & 13.748 $\pm$ 0.007 & 13.095 $\pm$ 0.011 \\
15-11-2022 & 59898.60 & YAHPT & 15.291 $\pm$ 0.008 & 14.385 $\pm$ 0.007 & 13.758 $\pm$ 0.007 & 13.116 $\pm$ 0.010 \\
17-11-2022 & 59900.60 & YAHPT & 15.292 $\pm$ 0.008 & 14.384 $\pm$ 0.007 & 13.742 $\pm$ 0.007 & 13.121 $\pm$ 0.010 \\
20-11-2022 & 59903.50 & YAHPT & 15.303 $\pm$ 0.006 & 14.411 $\pm$ 0.006 & 13.783 $\pm$ 0.006 & 13.164 $\pm$ 0.009 \\
21-11-2022 & 59904.49 & YAHPT & 15.306 $\pm$ 0.006 & 14.416 $\pm$ 0.006 & 13.786 $\pm$ 0.006 & 13.167 $\pm$ 0.009 \\
24-11-2022 & 59907.50 & YAHPT & 15.316 $\pm$ 0.007 & 14.420 $\pm$ 0.006 & 13.792 $\pm$ 0.006 & 13.168 $\pm$ 0.009 \\
25-11-2022 & 59908.49 & YAHPT & 15.310 $\pm$ 0.006 & 14.422 $\pm$ 0.006 & 13.793 $\pm$ 0.006 & 13.165 $\pm$ 0.009 \\
27-11-2022 & 59910.50 & YAHPT & 15.308 $\pm$ 0.007 & 14.420 $\pm$ 0.006 & 13.784 $\pm$ 0.006 & 13.177 $\pm$ 0.009 \\
28-11-2022 & 59911.49 & YAHPT & 15.316 $\pm$ 0.007 & 14.424 $\pm$ 0.006 & 13.793 $\pm$ 0.006 & 13.161 $\pm$ 0.009 \\
01-12-2022 & 59914.49 & YAHPT & 15.321 $\pm$ 0.007 & 14.438 $\pm$ 0.006 & 13.807 $\pm$ 0.006 & 13.199 $\pm$ 0.009 \\
02-12-2022 & 59915.49 & YAHPT & 15.328 $\pm$ 0.007 & 14.440 $\pm$ 0.006 & 13.815 $\pm$ 0.006 & 13.207 $\pm$ 0.010 \\
03-12-2022 & 59916.56 & YAHPT & 15.301 $\pm$ 0.012 & 14.402 $\pm$ 0.008 & 13.782 $\pm$ 0.007 & 13.178 $\pm$ 0.011 \\
05-12-2022 & 59918.49 & YAHPT & 15.313 $\pm$ 0.008 & 14.440 $\pm$ 0.007 & 13.819 $\pm$ 0.007 & 13.191 $\pm$ 0.010 \\
06-12-2022 & 59919.48 & YAHPT & 15.304 $\pm$ 0.008 & 14.427 $\pm$ 0.007 & 13.616 $\pm$ 0.083 & 13.209 $\pm$ 0.010 \\
07-12-2022 & 59920.52 & YAHPT & 15.314 $\pm$ 0.011 & 14.420 $\pm$ 0.007 & 13.815 $\pm$ 0.007 & 13.210 $\pm$ 0.011 \\
08-12-2022 & 59921.52 & YAHPT & 15.297 $\pm$ 0.012 & 14.421 $\pm$ 0.009 & 13.825 $\pm$ 0.008 & 13.230 $\pm$ 0.014 \\
09-12-2022 & 59922.49 & YAHPT & 15.319 $\pm$ 0.008 & 14.434 $\pm$ 0.007 & 13.812 $\pm$ 0.007 & 13.201 $\pm$ 0.010 \\       
\hline
\end{tabular}
\end{table*}

\end{appendix}

\end{document}